\documentclass[iop, numberedappendix,apj]{emulateapj}

\usepackage{natbib}

\shorttitle{Gemini Planet Imager Polarimetry of HR 4796A}
\shortauthors{Perrin et al.}

\usepackage{color}
\definecolor{blue}{rgb}{0.1,0.1,0.6}
\definecolor{orange}{rgb}{0.74,.35,0.099}
\definecolor{pale}{rgb}{0.90,0.90,0.95}
\definecolor{red}{rgb}{1.0,0.0,0.0}

\usepackage{xspace}
\newcommand{\betapic}{$\beta$~Pic\xspace}

\begin{document}

\submitted{Submitted to the Astrophysical Journal, 2014 July 2}
\title{Polarimetry with the Gemini Planet Imager:  Methods, Performance at First Light, and the Circumstellar Ring around HR 4796A}

\shorttitle{Polarimetry with GPI and the Ring around HR 4796A}

\author{Marshall D. Perrin\altaffilmark{1}, 
Gaspard Duchene\altaffilmark{2,3,4}
Max Millar-Blanchaer\altaffilmark{5}, 
Michael P. Fitzgerald\altaffilmark{6}, 
James~R.~Graham\altaffilmark{2}, 
Sloane J. Wiktorowicz\altaffilmark{7,8},
Paul G. Kalas\altaffilmark{2}, 
Bruce Macintosh \altaffilmark{9,10}, 
{Brian Bauman}\altaffilmark{9},
{Andrew Cardwell}\altaffilmark{11},
{Jeffrey Chilcote}\altaffilmark{6},
{Robert J. De Rosa}\altaffilmark{12,13},
{Daren Dillon}\altaffilmark{7},
{Ren\'e Doyon}\altaffilmark{14},
{Jennifer Dunn}\altaffilmark{15},
{Donald Gavel}\altaffilmark{7},
{Stephen Goodsell}\altaffilmark{11},
{Markus Hartung}\altaffilmark{11},
{Pascale Hibon}\altaffilmark{11},
{Patrick Ingraham}\altaffilmark{10,14},
Daniel Kerley\altaffilmark{15},
{Quinn Konapacky}\altaffilmark{5},
{James E. Larkin}\altaffilmark{6},
{J\'er\^ome Maire}\altaffilmark{5},
{Franck Marchis}\altaffilmark{16},
{Christian Marois}\altaffilmark{15},
Tushar Mittal\altaffilmark{2},
{Katie M. Morzinski}\altaffilmark{17,8},
{B.~R.~Oppenheimer}\altaffilmark{18},
{David W. Palmer}\altaffilmark{9},
{Jennifer Patience}\altaffilmark{12},
{Lisa Poyneer}\altaffilmark{9},
{Laurent Pueyo}\altaffilmark{1},
{Fredrik T. Rantakyr\"o}\altaffilmark{11},
{Naru Sadakuni}\altaffilmark{11},
{Leslie Saddlemyer}\altaffilmark{15},
{Dmitry Savransky}\altaffilmark{19},
{R\'emi Soummer}\altaffilmark{1},
{Anand Sivaramakrishnan}\altaffilmark{1},
{Inseok Song}\altaffilmark{20},
{Sandrine Thomas}\altaffilmark{21},
{J. Kent Wallace}\altaffilmark{22},
{Jason J. Wang}\altaffilmark{2},
{Schuyler G. Wolff}\altaffilmark{23,1}
}

\altaffiltext{1}{Space Telescope Science Institute}
\altaffiltext{2}{Astronomy Department, University of California, Berkeley}
\altaffiltext{3}{Univ. Grenoble Alpes, IPAG, F-38000 Grenoble, France}
\altaffiltext{4}{CNRS, IPAG, F-38000 Grenoble, France}
\altaffiltext{5}{Dept. of Astronomy \& Astrophysics, University of Toronto}
\altaffiltext{6}{Dept. of Physics and Astronomy, UCLA}
\altaffiltext{7}{Dept. of Astronomy, UC Santa Cruz}
\altaffiltext{8}{NASA Sagan Fellow}
\altaffiltext{9}{Lawrence Livermore National Laboratory}
\altaffiltext{10}{Stanford University}
\altaffiltext{11}{Gemini Observatory, La Serena, Chile}
\altaffiltext{12}{Arizona State University}
\altaffiltext{13}{School of Physics, University of Exeter, Exeter, UK}
\altaffiltext{14}{Department de Physique, Universit\'{e} de Montr{\'e}al}
\altaffiltext{15}{National Research Council of Canada Herzberg}
\altaffiltext{16}{SETI Institute, Mountain View, CA }
\altaffiltext{17}{Steward Observatory, University of Arizona}
\altaffiltext{18}{American Museum of Natural History, New York}
\altaffiltext{19}{Cornell University}
\altaffiltext{20}{Dept. of Physics and Astronomy, University of Georgia, }
\altaffiltext{21}{NASA Ames Research Center,  }
\altaffiltext{22}{NASA Jet Propulsion Laboratory}
\altaffiltext{23}{Physics \& Astronomy Dept., Johns Hopkins University}

\begin{abstract}
We present the 
first results from the polarimetry mode of the Gemini Planet Imager (GPI), which uses a new integral field polarimetry architecture to provide high contrast linear polarimetry with minimal systematic biases between the orthogonal polarizations. We describe the design, data reduction methods, and performance of polarimetry with GPI. Point spread function subtraction via differential polarimetry suppresses unpolarized starlight by a factor of over 100, and provides sensitivity to circumstellar dust reaching the photon noise limit for these observations.
In the case of the circumstellar disk around HR 4796A, 
GPI's advanced adaptive optics system reveals the disk clearly even prior to PSF subtraction. In polarized light, the disk is seen all the way in to its semi-minor axis for the first time. The disk exhibits surprisingly strong asymmetry in polarized intensity, with the west side $\gtrsim 9$ times brighter than the east side despite the fact that the east side is slightly brighter in total intensity. Based on a synthesis of the total and polarized intensities, we now believe that the west side is closer to us, contrary to most prior interpretations. Forward scattering by relatively large silicate dust particles leads to the strong polarized intensity on the west side, and the ring must be slightly optically thick in order to explain the lower brightness in total intensity there. 
These findings suggest that the ring is geometrically narrow and dynamically cold, perhaps shepherded by larger bodies in the same manner as Saturn's F ring.
\end{abstract}


\section{Introduction}

\setcounter{footnote}{0}

In the thirty years since the pioneering coronagraphy of the disk around \betapic \citep{SmithTerrile1984}, imaging observations of nearby planetary systems have blossomed into a rich field of study. Young pre-main-sequence stars are now commonly seen to have optically thick gas- and dust-rich protoplanetary disks, while more tenuous debris disks have been spatially resolved around dozens of main-sequence stars \citep[see][and references therein.]{Williams2011ARAA,Matthews2014PP6}. 
Disks are intimately related to planets, both giving rise to planets in young protoplanetary disks and in turn arising as second-generation debris disks from the collisional destruction of planetesimals. Indeed, \betapic itself is now seen to have a Jovian planet orbiting within and sculpting its disk \citep{Lagrange2009,Lagrange2010}, and a correlation between the presence of planets and debris disks is increasingly supported by observational data \citep{Raymond2012,Bryden2013AAS,Marshall2014}. Imaging studies complement spectroscopy and interferometery, and remain essential for investigating many aspects of the physics of circumstellar disks. 
Each new generation of instruments offers further clarity on the details of how planetary systems form and evolve.

The key challenge in imaging disks at optical and near-infrared wavelengths is achieving sufficient contrast to detect faint and extended disk-scattered light in the presence of a brighter stellar point spread function (PSF). 
While a coronagraph can help block direct starlight, particularly if that light is concentrated in a coherent diffraction-limited PSF, coronagraphs inevitably still leave some unblocked residual starlight. Differential measurement techniques and PSF subtraction in software are needed to remove this remnant starlight to provide a clear view of circumstellar disks.

Differential imaging polarimetry and adaptive optics (AO) have proven to be a particularly useful combination toward this goal.   Randomly polarized starlight assumes a preferential polarization state after scattering off circumstellar dust; the degree of polarization and its dependence on scattering angle and wavelength are set by the nature of the scattering particles.  Not only can the induced polarization be used as a diagnostic for the properties of those particles \cite[e.g.][]{Graham2007ApJ...654..595G}, it also provides a powerful discriminant of disk-scattered light from the intrinsically unpolarized stellar PSF \citep[e.g.][]{Kuhn2001ApJ...553L.189K}.  By measuring orthogonal polarizations simultaneously, the residual star light common to the two polarizations should vanish in their difference; modulation of the sensed polarization state (e.g. with a half wave plate) can further mitigate residual instrumental non-common-path artifacts. PSF suppression of unpolarized starlight by factors of 50--100 can typically be achieved \citep{Perrin2008ircalpol, Hinkley09a}, allowing sensitive detections of circumstellar dust even in cases of only relatively modest AO image quality (for example, many images of disks have been obtained in AO observations with Strehl ratios as low as 0.1-0.2;  \citealp{Perrin2004Sci...303.1345P, Hashimoto2012, Follette2013}). Differential polarimetry instruments have been deployed at most large telescopes with AO, including Gemini North, the Lick 3 m, Subaru, the Very Large Telescope, and others. These have yielded observations of both debris and protoplanetary disks. See \citet{Perrin2014polbook} for a recent review. 

Building upon these successes,  imaging polarimetry capabilities have been incorporated into the latest generation of AO systems optimized for high contrast, including the Gemini Planet Imager (GPI; \citealp{Macintosh2008,Macintosh2014}) and SPHERE \citep{Beuzit2008,Langlois2010SPIE.7735E..97L,Schmid2010lyot.confE..49H}.  GPI is a dedicated system for directly imaging and spectroscopically characterizing extrasolar planets. It combines very high order AO \citep{Poyneer2008,Poyneer2014}, a diffraction-suppressing coronagraph \citep{Soummer2011}, an integral field spectrograph with low spectral resolution but high spatial resolution \citep{Chilcote2012SPIE,Larkin2014}, and an interferometric calibration wavefront sensor unit, plus an optomechanical support structure and extensive electronics and software infrastructure.
First light of GPI occurred on 2013 November 12 UT; its initial performance achieved sufficient contrast for detection of \betapic~b in single raw 60 s exposures without any postprocessing. Details on instrument design, first light performance including AO system wavefront error budgets, and initial exoplanetary science results are reported in \cite{Macintosh2014}.

For the study of circumstellar disks, GPI has a science goal of detecting debris disks with an optical depth of $\geq 4 \times 10^{-5}$, corresponding to the majority of disks detected by infrared excesses in the {\it IRAS} catalog \citep{Zuckerman2004}. 
To achieve this, GPI's AO system and coronagraph are augmented with a polarimeter that suppresses speckles of unpolarized starlight by factors of $>100$. In this paper we describe the design and performance of GPI's polarimetry mode and present its first science results, obtained as part of the GPI on-sky verification and commissioning program.  The first polarimetry observations with GPI were made on 2013 December 12 UT. GPI polarimetry has since been extensively tested and characterized in subsequent commissioning runs, and polarimetry made up approximately half of the shared-risk early science observations in April 2014. 

The first half of this paper describes the properties and performance of GPI as a polarimeter, starting with the optical design and hardware implementation  (\S \ref{design_and_implementation} and Appendix A). We describe the data reduction techniques we have developed (\S \ref{data_reduction} and Appendix B) and made available to the community as part of our open-source GPI Data Reduction Pipeline \citep{Perrin2014}.  We evaluate performance at first light (\S \ref{performance_tests}) using observations of standard stars to assess
achieved starlight suppression and polarimetric precision and accuracy. 

In the second half of this paper, to demonstrate GPI's capabilities for disk imaging we present observations of the circumstellar ring around HR 4796A. GPI observations provide high contrast images of the disk in both total
and polarized intensity (\S \ref{hr4796a}) that reveal a strong asymmetry in the degree of polarization, surprising given the relatively isotropic total intensity phase function. Based on our analysis of these observations, we now believe this ring to be optically thick and dominated by scattering from relatively large dust particles (\S \ref{sec:analysis}).  All data presented in this paper are from the December 2013 and March 2014 GPI observing runs. 
Subsequent multi-wavelength followup observations of HR 4796A will be presented in a future paper (Fitzgerald et al. 2014, in prep). We conclude with some brief remarks on future prospects for circumstellar disk studies with GPI, SPHERE, and other high contrast systems (\S 7).

\section{The Imaging Polarimetry Mode for GPI}
\label{design_and_implementation}

The design of GPI's polarimetry mode was previously presented by \citet{Perrin2010SPIE}, and builds upon concepts and lessons learned from earlier AO polarimeters \citep[as summarized in][]{Perrin2008ircalpol}.
The imaging polarimetry mode is implemented primarily by a polarizing Wollaston prism beamsplitter within the IFS \citep{Larkin2014} and a rotating half-wave plate within the calibration unit, but depends on the overall instrument system to achieve its performance goals.

\begin{figure*}
\begin{center}
\includegraphics[width=6.5in]{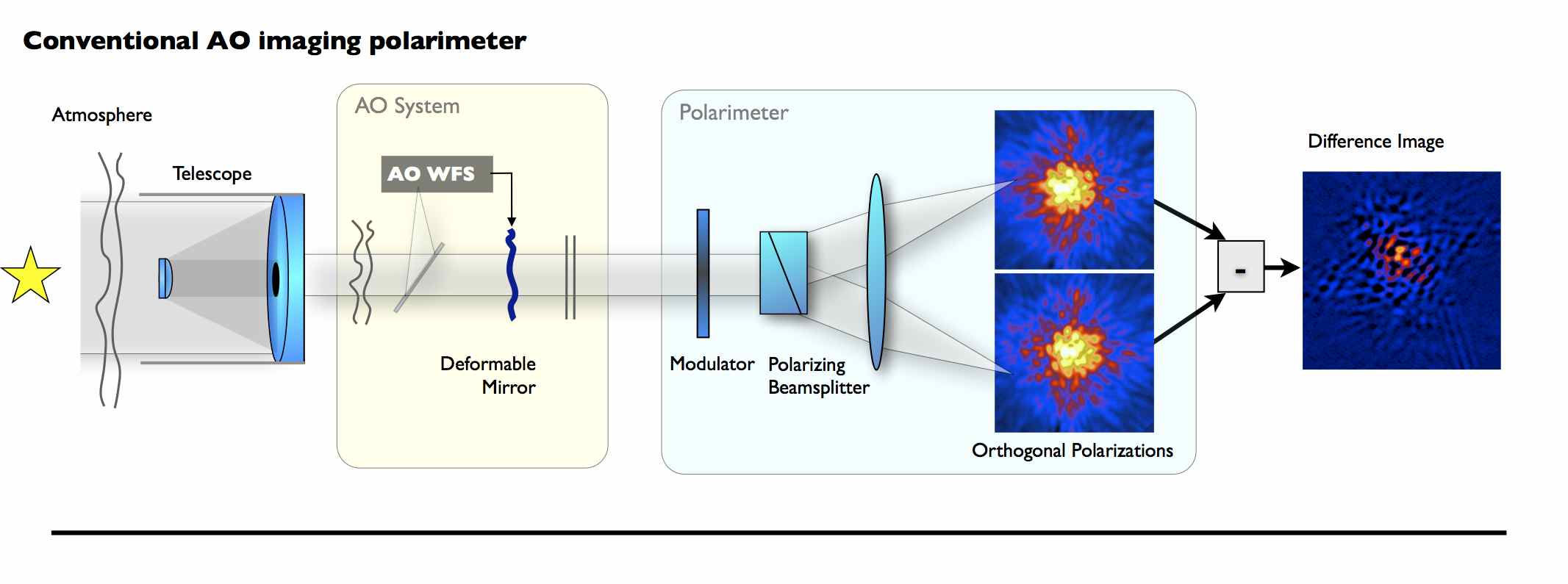}
\includegraphics[width=6.5in]{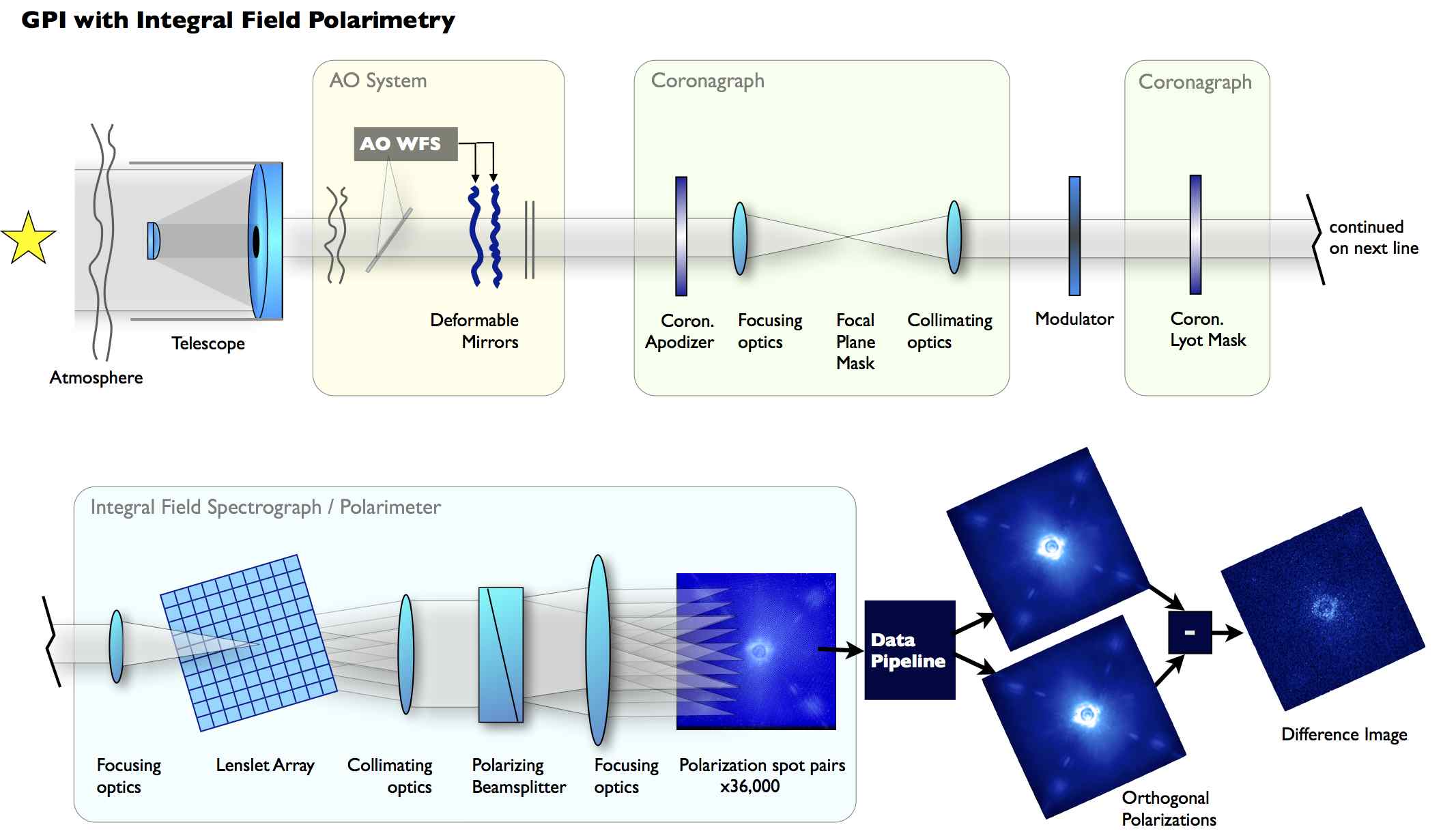}
\end{center}
\caption{ Conceptual cartoons of a traditional AO polarimeter and the integral field polarimetry architecture
    developed for GPI, indicating the order of the various optical components. These figures are highly simplified, with many optics omitted and others shown as transmissive even though in practice most are reflective. 
\textit{Upper panel:}   In a conventional AO polarimeter, the orthogonally polarized PSFs are similar but not identical due to non-common-path wavefront errors and the wavelength dependence of the Wollaston prism's birefringence; modulation with a rotating waveplate mitigates these differences. The data shown here for the conventional AO polarimeter are observed PSFs from the Lyot Project coronagraph at the AEOS 3.6 m \citep{Hinkley09a}.
\textit{Lower panel:}
 In an integral field polarimeter, the light is optically divided by the lenslet array prior to any dispersion, largely eliminating sensitivity to non-common-path errors and chromatic birefringence and thus producing a difference image with speckles highly suppressed. 
}
\label{hardware_cartoons}
\end{figure*}

\subsection{Design Principles}

Our central criterion when designing GPI's polarimetry mode was to optimize contrast in spatially resolved coronagraphy of circumstellar disks. This led to several design choices made differently than for a typical point-source polarimeter, and differently from what we would have adopted if our goal had been the utmost performance in absolute polarimetric accuracy.   GPI is designed first and foremost as an integrated starlight suppression system, only secondarily as a polarimeter.

In particular, we sought to minimize static wavefront errors prior to the coronagraphic focal plane.  This is necessary because coronagraphic starlight suppression depends sensitively on image quality at the coronagraph's focal plane---aberrations after that point have much less ability to create speckles since most starlight has been removed.  The conventional wisdom for high precision polarimetry is that a modulator (e.g.\ waveplate) should be placed as early as possible in the optical path, to mitigate instrumental polarization.  However for GPI this would have meant one or more additional pre-coronagraph optical surfaces, likely with relatively poor wavefront quality\footnote{It is impractical to fabricate waveplates better than $\sim 50$ nm peak-to-valley transmitted wavefront error, since repeated polishing to improve surface quality is incompatible with controlling the absolute thickness in order to set the retardance. Alternate modulators such as liquid crystal retarders tend to have even worse wavefront quality than waveplates, $\geq 100$ nm RMS. For comparison the superpolished reflective optics used in GPI typically have $< 1$ nm rms.}. We therefore decided to locate an achromatic waveplate retarder \textit{after} the coronagraphic focal plane,  specifically located in a collimated beam in the input relay to the IFS and mounted on a linear stage so that it can be removed from the light path during spectral-mode observations.

This design means that we cannot modulate out the instrumental polarization induced by off-axis reflections from GPI's many optics, but must instead calibrate and remove this bias during data reduction (\S \ref{sec:instr_pol}).  However this approach was inevitable for GPI, since Gemini instruments are required to operate at any port of the Gemini Instrument Support Structure, most of which are fed by a 45\degr\ reflection off the telescope's tertiary mirror. This tertiary fold is likely a dominant contributor to instrumental polarization when mounted on a side port, and its effects could not possibly be mitigated by any modulator within GPI\footnote{Facility-class polarization modulators for each Gemini telescope were designed and built to avoid precisely this problem, but these GPOL systems remain uncommissioned and are unlikely to ever be used.}. 
Instrumental polarization from a 45\degr\ fold can be cancelled optically by balanced reflections from a second mirror tilted on an axis orthogonal to that of the first mirror \citep{Cox1976}, but we did not pursue that approach because it would do more harm than good for the case in which GPI is mounted on the straight-through bottom port. 
Thus far GPI has only been operated at that bottom port, so currently the tertiary fold does not contribute to instrumental polarization, and as of mid-2014 it is planned that GPI will remain on the bottom port for future observing runs.

\subsection{Integral Field Polarimetry}
\label{section:ifp}

\begin{figure*}[Ht!]
\begin{center}
\includegraphics[height=2in]{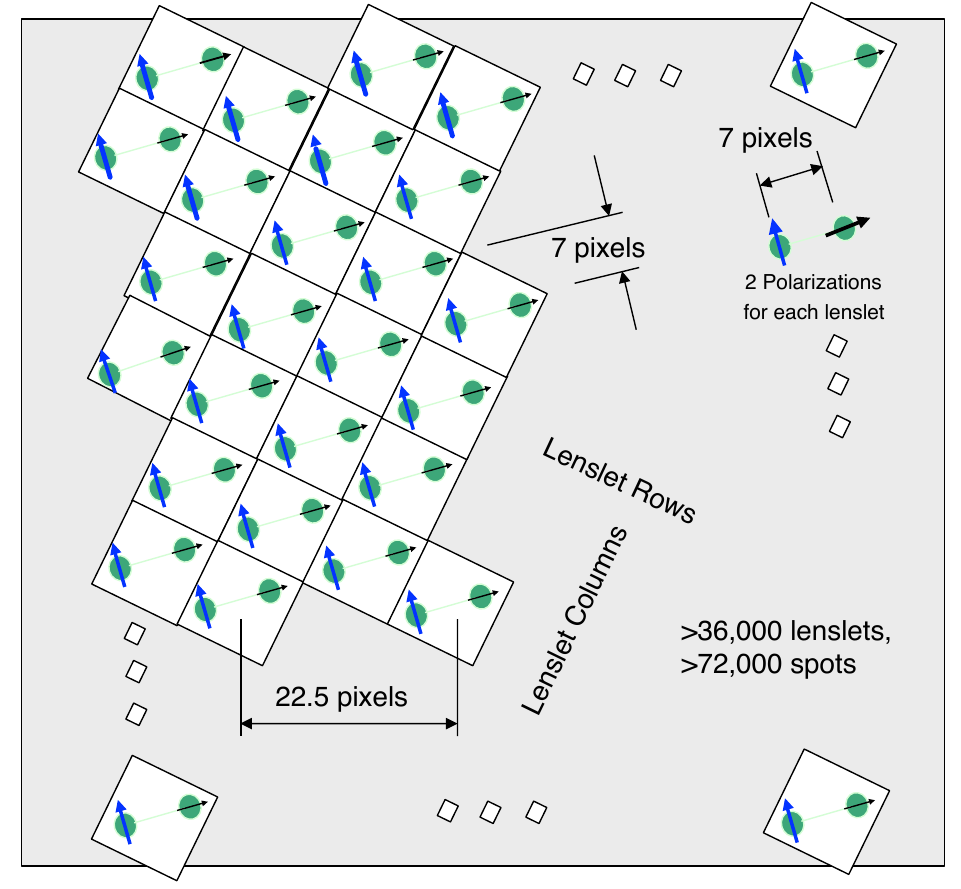} 
\includegraphics[height=2in]{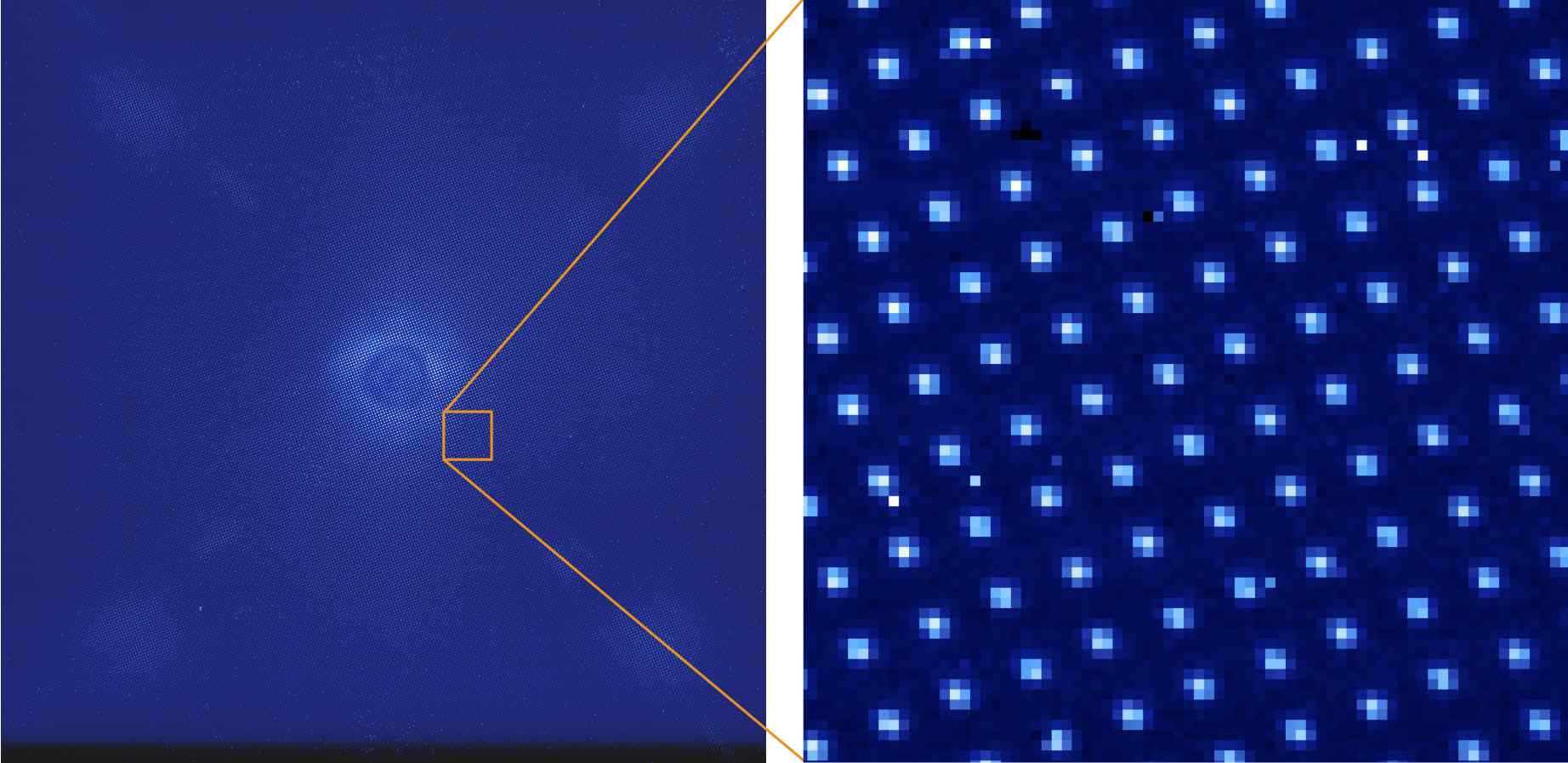} 
\end{center}
\caption{ \textit{Left:} Schematic of polarized spot layout for integral field polarimetry. Spots are dispersed at 45\degr\ relative to the lenslet grid in order to maximize the spacing between adjacent spots of orthogonal polarization.  \textit{Center:} Actual raw GPI IFS image in polarization mode, showing the full spot array. The dark hole created by the AO system and coronagraph is faintly visible, along with the square of satellite spots used for astrometric and photometric calibration. \textit{Right:} A zoomed-in panel showing a closer view of some of the polarized spots.  Alternating spots in the manner of a checkerboard correspond to the same linear polarization; the GPI data pipeline sums the flux in each spot and reassembles those values into a datacube giving two orthogonal linear polarizations. 
}
\label{spots-layout}
\end{figure*}

Given that the science instrument of GPI is a lenslet-based IFS, we implement the measurement of polarized light via a novel mode within that IFS that we call ``integral field polarimetry''.  In this mode the regular spectral dispersing prism within the IFS is replaced with a Wollaston prism, giving up spectral resolution to gain sensitivity to orthogonal linear polarizations. This swap is implemented by mounting both prisms on a linear stage within the cryogenic volume of the IFS.  The polarizations are thus dispersed only \textit{after} the lenslet array has pixelated the field of view. Each lenslet in the array corresponds to one spatial resolution element or spaxel. Figure \ref{hardware_cartoons} depicts this architecture in comparison with traditional AO polarimeter.

Integral field polarimetry minimizes differential wavefront error between the two polarizations, eliminates the difficulty of precise image registration that has limited previous AO imaging polarimeters \citep[e.g.][]{Perrin2004Sci...303.1345P,Apai2004A&A...415..671A}, and does not require any reduction in field of view. It also prevents any distortion of PSFs due to lateral chromatism of the Wollaston prism, thus removing any need for exotic materials with low birefringent chromaticity such as YLF \citep{Oliva1997A&AS..123..179O,Perrin2008ircalpol}.

The optimal Wollaston prism design for integral field polarimetry maximizes the separation of the polarized spots by dispersing each lenslet along a $\pm 45$\degr\ axis relative to the lenslet grid. This results in two sets of dots interlaced in the same manner as the black and white squares of a chessboard.  See Figure \ref{spots-layout}. In addition to maximally separating the spots, this design results in the diffraction pattern from the square lenslet grid (the dominant form of crosstalk between spots) only overlapping onto adjacent spots of the same polarization, thus minimizing contamination between channels.  

Further details of the polarizing prism, waveplate, and associated mechanisms for GPI are given in Appendix \ref{appendix:details}.

\section{Data Reduction Methods}
\label{data_reduction}

\subsection{Key Concepts}

Polarimetric observations with GPI consist of a series of exposures taken with the  Wollaston and waveplate inserted into the beam, with the waveplate rotating between each subsequent exposure. The waveplate rotation angle is a free parameter that in principle can be set arbitrarily for each exposure; in practice, we generally follow the typical approach of stepping 22.5$^\circ$ per move: 0, 22.5, 45, 67.5, and so on\footnote{During much of commissioning, a minor software bug in the Gemini Observing Tool (OT) restricted the waveplate position to integers, so we used 22\degr\ and 68\degr\ instead of 22.5\degr\ and 67.5\degr. The data reduction pipeline handles any arbitrary sequence of rotation angles. The OT was subsequently updated to allow floating point values. See Appendix A for more details on the waveplate modulation mechanism.}. These images must then be combined to extract the astronomical polarization signal.

GPI always remains fixed with respect to the telescope pupil, so the sky appears to constantly rotate,  allowing angular differential imaging (ADI; \citealp{Marois2006ApJ}) to further reduce the residual speckle halo. The fact that GPI always observes in ADI mode complicates polarimetric data reduction because each exposure has unique sky-projected polarization axes and field rotation. Traditional double-differencing methods of polarization data reduction do not suffice. Instead, we combine each sequence of exposures based on a more generalized forward model of observations: for each individual exposure, we construct the particular Mueller matrix equation describing how sky rotation and the polarizing optics within GPI map astrophysical polarizations into orthogonal polarization signals measured on the detector. By inverting the resulting set of equations through a least-squares analysis, we derive the astrophysical polarization signal for each position in the field of view. The result is a 3D Stokes datacube [$x$, $y$, ($I,Q,U,V$)] derived from the entire observation sequence. Further details of this algorithm are provided in Appendix \ref{appendix:leastsquares}.

One of the strengths of differential polarimetry is that modulation of the polarized signal can mitigate non-common-path biases between the orthogonal channels. For traditional imaging polarimeters, this cancellation can be obtained simply by creating double differences of images appropriately modulated to swap opposite Stokes parameters such as $+Q$ and $-Q$ \citep[e.g.][]{Kuhn2001ApJ...553L.189K}. This, too, is complicated by the fact that GPI observations are always ADI mode.  Such images cannot be added together until after a derotation step brings them to common orientation in the sky frame, but successive derotated images no longer align in the instrument frame so instrumental non-common-path biases will not cancel. Integral field polarimetry does mostly eliminate the potential for any optical non-common-path biases (i.e., light is not dispersed in polarization until after it has been optically pixelated). Detector artifacts and imperfect datacube assembly can still induce non-common-path offsets between the two channels for which we must compensate. In place of the traditional double-differencing, we have included an analysis step that estimates the bias between the two orthogonal polarization channels based on the mean channel difference for all the exposures in a given observing sequence. Subtracting this mean difference from each individual difference image yields a variant of double differenced image. Empirically this method greatly reduces systematics in the final datacubes for GPI. Details of this algorithm are described further in Appendix \ref{appendix:doubledifferencing}.

\subsection{ Steps of Data Processing}
\label{sec:dataprocessing}

\begin{figure*}[Ht!]
\begin{center}
\includegraphics[width=6in]{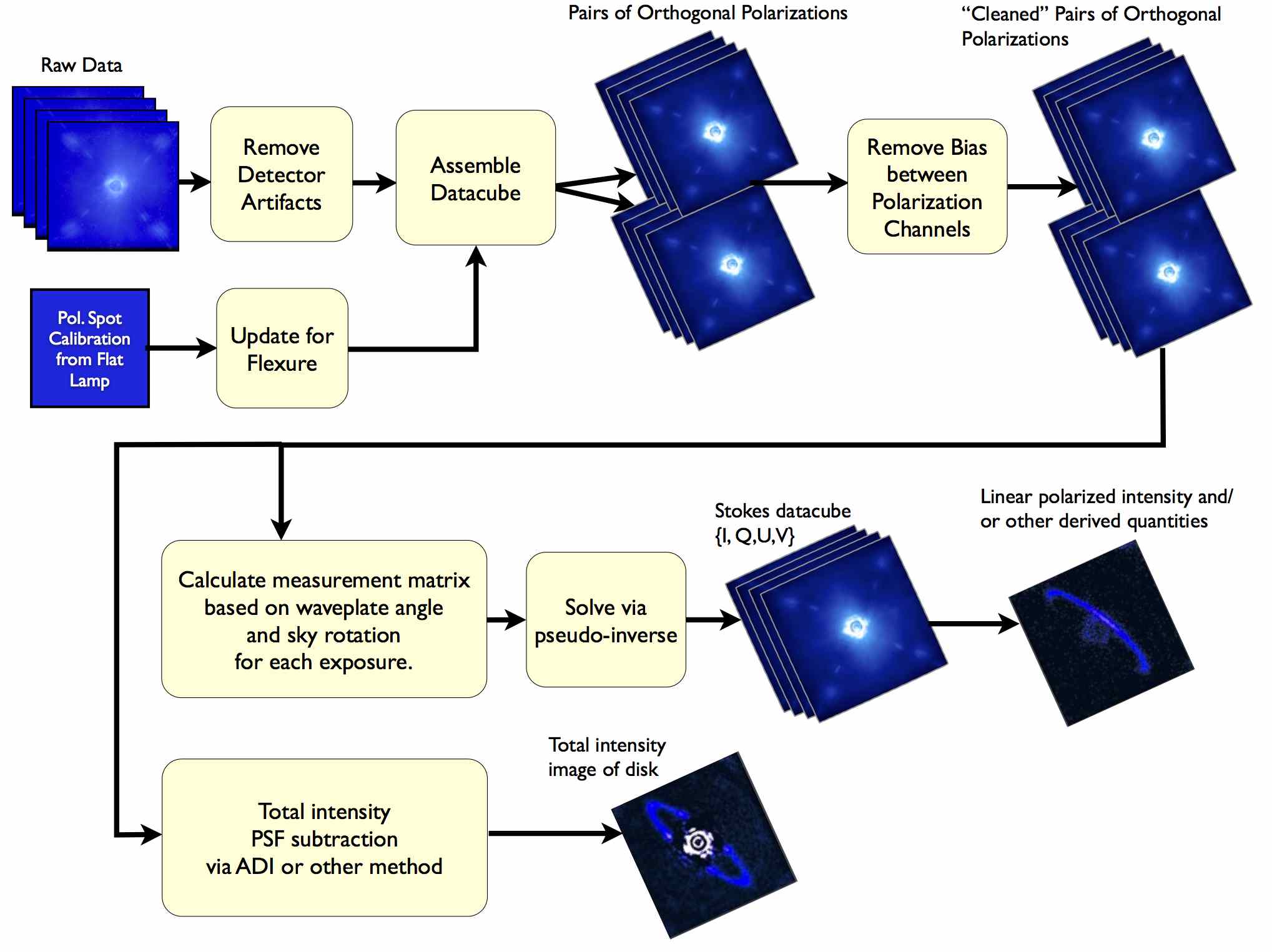} 
\end{center}
\caption{ \label{fig:dataprocessing} Sketch of key steps in reducing GPI polarimetry data, from raw data to polarized intensity of scattered light.  
See the text in section \ref{sec:dataprocessing} for description of each step.
}
\end{figure*}

The steps involved in processing GPI polarimetry are depicted in schematic form in Figure \ref{fig:dataprocessing}. Software implementing these tasks is available as part of the GPI Data Reduction Pipeline\footnote{Software and documentation available from http://planetimager.org/datapipeline/.} \citep{Perrin2014}. 
Starting from raw files written by the instrument, we first subtract a dark image in the usual manner, followed by a ``destriping" step to remove correlated noise due to the readout electronics and microphonics noise induced in the H2RG detector by mechanical vibration from the IFS cryocoolers \citep{Chilcote2012SPIE,Ingraham2014Detectors}. 
The impact of these noise terms decreases as $1/\sqrt{N_{reads}}$, and thus is most significant for short integration times (small numbers of up-the-ramp reads); for typical science exposures of a minute or longer, the impact of correlated noise is minimal and the destriping step is often unnecessary. 

To assemble a pair of polarization images from the raw data, we must know the locations of the tens of thousands of dispersed polarization spots. The spot pattern is close to but not perfectly uniform due to distortions in the spectrograph optics. A calibration map of all spot locations is made by fitting model lenslet PSFs to a high S/N polarimetry mode observation of the GCAL flat field quartz halogen lamp. 
Due to internal flexure in the spectrograph as a function of elevation, the spot locations during nighttime observations will be slightly shifted relative to daytime calibration flats (typically less than one pixel, and at most a few pixels). The offset is estimated based on a subset of the brightest spots and applied to all spot positions. 

Flux in each pol spot is summed over a small aperture. Currently, a fixed $5\times5$ pixel box is used, but future versions of the pipeline will perform optimal extraction using empirically measured microlens PSFs \citep{Ingraham2014MicrolensPSFs}; this will provide an improvement in S/N and increased robustness to bad pixels.  Iterating over the whole detector and repeating for both polarizations yields a pair of orthogonally polarized images.  A flat field correction for variations in lenslet throughput can then be applied. Residual bad pixels may be detected as local statistical outliers, and their values replaced through interpolation.
Using the entire set of observed images, the systematic bias between the orthogonal polarization channels is estimated and subtracted following the methods in Appendix \ref{appendix:doubledifferencing}.

For coronagraphy, the position of the occulted star is estimated via a Radon-transform-based algorithm \citep{Pueyo2014,Wang2014} using the satellite spots created by a diffractive grid on the GPI apodizers \citep{Sivaramakrishnan2006,Marois2006SatSpots}. Based on the derived centers, each exposure in the sequence is rotated to north up and aligned to a common position. 
The files are then combined via least squares following the methods outlined in \S \ref{appendix:leastsquares} to produce a Stokes datacube with slices $\{I,Q,U,V\}$.\footnote{GPI's sensitivity to circular polarization $V$ is small but nonzero since the waveplate retardance is not precisely 0.5 waves.}  A single Stokes datacube is the result of each observation sequence. Derived quantities such as the polarized intensity, $P = (Q^2+U^2)^{1/2}$, and polarization fraction, $p = P/I$, may then be calculated.

Differential polarimetry via the above steps yields increased contrast in polarized light, but does not do so for total intensity. If a high contrast image in total intensity is desired, it must be obtained via other methods, for instance PSF subtractions using ADI. Methods such as LOCI \citep{Lafreniere2007} or KLIP \citep{Soummer2012ApJ} can be effective, but require observers to carefully model the effects of these algorithms on the disk's apparent surface brightness and morphology \citep[e.g.,][]{Esposito2014ApJ...780...25E,Rodigas2014ApJ...783...21R,Milli2014arXiv1405.2536M}.

\section{Performance}

\label{performance_tests}

\subsection{Suppression of Unpolarized Starlight}
\label{sec:suppression}
As part of the GPI commissioning program, we have measured the polarimetric suppression of residual starlight based on observations of unpolarized standard stars.  On 2014 March 24 the unpolarized standard HD 118666 was observed with GPI in closed-loop coronagraph mode with the $J$, $H$ and $K1$ filters (see Table~\ref{table:observations}). HD 118666 is a bright ($H$=4.8) nearby F3III-IV star with a measured polarization of $0.07\% \pm 0.035$ \citep{Mathewson70}.

\begin{figure}[ht!]
\begin{center}
\includegraphics[width=3.5in]{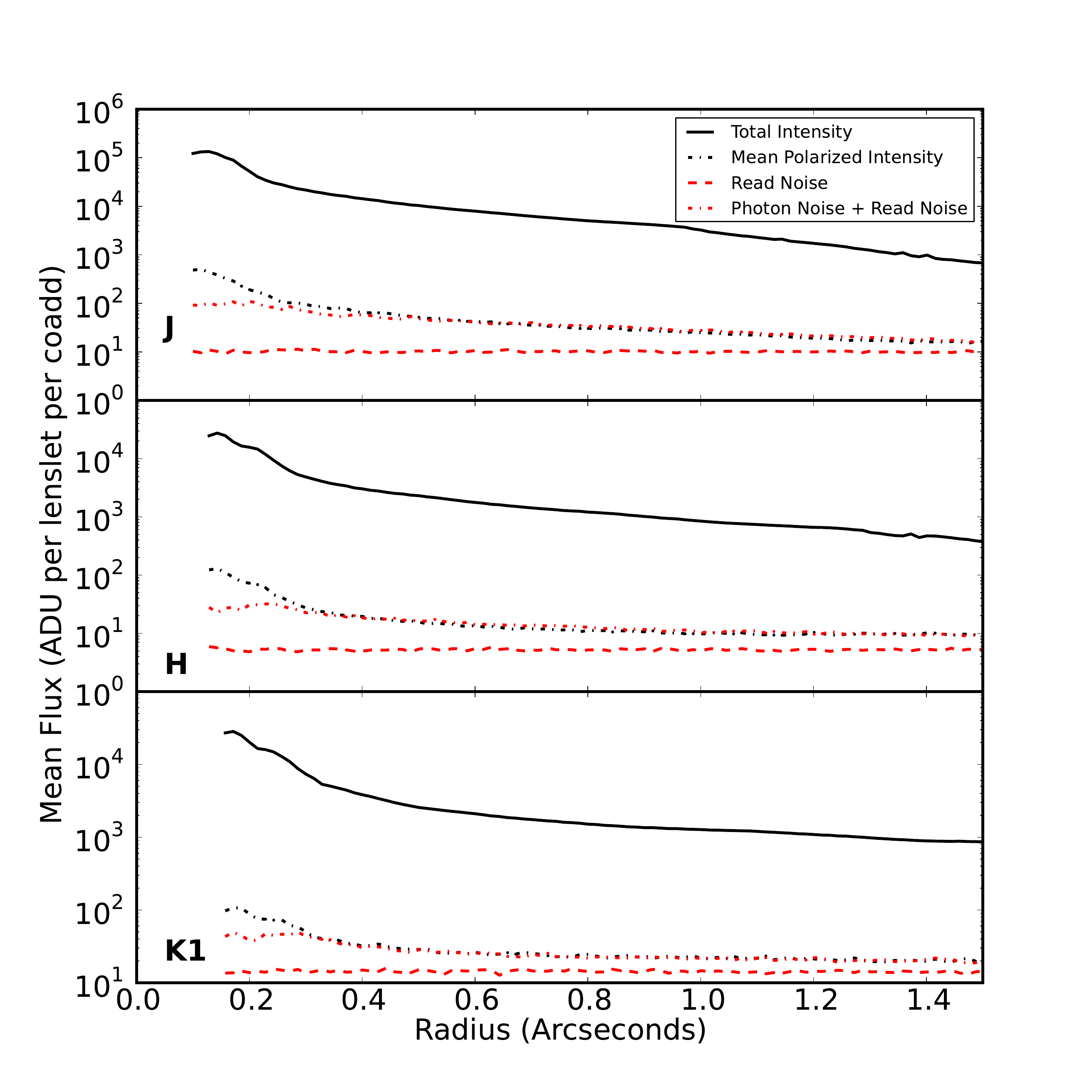} 
\end{center}
\caption{
Total intensity and linear polarized intensity as a function of radii for observations of the unpolarized standard star HD 118666 in GPI's $J$, $H$ and $K1$ bands. The read noise and photon noise curves were calculated with simulated data and indicated that these observations are dominated in all three bands by photon noise at radii greater than $\sim$0.4".
}
\label{fig:speckle_suppression_profiles}
\end{figure}

\begin{figure}[hbt!]
\begin{center}
\includegraphics[width=3.5in]{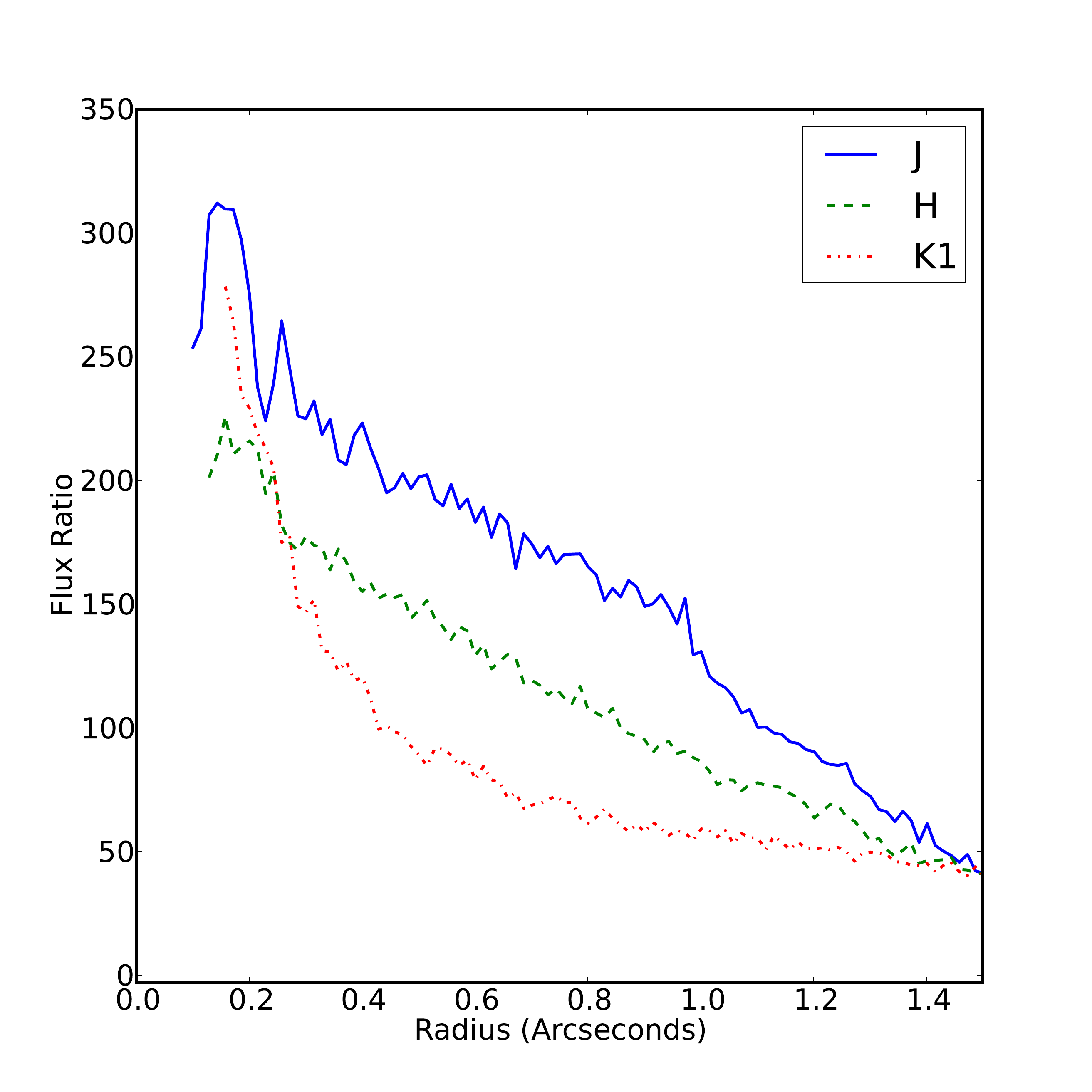} 
\end{center}
\caption{
The ratio of the total intensity to the linear polarized intensity. At angular separations smaller than $\sim 0.2"$ unpolarized speckles are being suppressed by over a factor of 200. This value represents a lower limit on the performance of GPI, as instrumental polarization was not compensated for in this analysis. }
\label{fig:speckle_suppression_ratios}
\end{figure}

Each set of observations was reduced as described in the previous section to produce a Stokes data cube. We then measured, in annuli of increasing radius from the star, the mean total intensity (Stokes $I$) and the mean polarized intensity ($P$). The resulting radial profiles are shown in Figure~\ref{fig:speckle_suppression_profiles}. The ratio of these two quantities, $I/P$ as displayed in Figure~\ref{fig:speckle_suppression_ratios}, is a first-order measure of the suppression of unpolarized starlight. This quantity is formally the inverse of the apparent polarization fraction, but in this context is better considered as the ratio of pre- to post-differential-polarimetry total flux since the source is known to be unpolarized. In all bands, the total intensity is suppressed by factors of $>100$ at radii smaller than 0.4''. The suppression ratio increases to $>200$ inside 0.2'' and decreases at wider separations.

It is well known that polarized intensity is a positive definite quantity and thus biased upwards by any noise \citep[e.g.,][]{Vaillancourt2006}. To assess the contributions of different noise terms, we created simulated sets of observations containing only photon and read noise, repeated for each filter with the same number of exposures as the original datasets. 
The simulated orthogonal polarization images were constructed to have identical mean intensity radial profiles as the observations, and appropriate photon noise was added using the nominal gain of 3 e- ADU$^{-1}$. For read noise, we used prior measurements for the read noise at the relevant exposure times. 
The simulated datasets were then run through the reduction process to produce mock Stokes cubes, and the total intensity and polarized intensity were measured as a function of radius. Simulated sets of data that only included read noise, but not photon noise, were also created and reduced in the same fashion.\footnote{Note that the effective noise in output datacubes can also be estimated analytically as described in Appendix \ref{appendix:noise}. To do so accurately, one must properly consider the effects on the variance of all pipeline processing steps. In particular the ``Rotate North Up'' step, because it is an interpolation, acts to partially smooth the per-lenslet noise and introduce correlations between adjacent lenslets.  Empirically this is about a $\sim 20\%$ effect for the datasets considered here. Propagating mock data through the full pipeline naturally takes into account all such effects. }

The resulting mock polarized intensity profiles agree closely with the observations, confirming that photon noise is the dominant limiting factor at most separations.  Fig.~\ref{fig:speckle_suppression_profiles} plots the $P$  profiles from the simulated datasets in red, showing the good agreement with the observed $P$ profiles in black.  In all cases, the GPI data appears to be photon noise dominated at radii greater than 0.4", though the read noise is a significant secondary contributor. GPI's differential polarimetry mode achieves sensitivities to polarized circumstellar material reaching the fundamental photon noise limit set by the post-coronagraphic PSF.

At small radii, 0.15'' $< r <$ 0.3'', the observed polarized intensities rise above the photon noise floor for all filters.  The maximum starlight suppression factors achieved, 200 to 300, do not necessarily indicate any fundamental limitation of GPI. In particular, these measurements were made \textit{without} compensation for instrumental polarization; any nonzero instrumental polarization will directly contribute to the residual polarized intensity measured.  The instrumental polarization of $\sim$0.4\%, as measured below in \S \ref{sec:instr_pol}, is of precisely the right magnitude to explain the observed maximum suppression ratios. Greater suppression of starlight may be possible once the instrumental polarization calibration is finalized and incorporated into the reduction process.

\subsection{Polarized Twilight Sky Tests}

To validate the measurement of polarization angles it is helpful to observe a source of known polarization position angle (PA). The twilight sky provides a useful calibration target since it is highly polarized, predominantly due to Rayleigh scattering in the upper atmosphere. Other contributors to sky polarization include aerosol scattering, multiple scattering effects, and secondary illumination from light scattered off clouds, water, and land \citep[][and references therein]{Harrington2011}. In the case of pure Rayleigh scattering, which we consider here, the PA of the observed polarization is aligned 90 degrees from the solar azimuth (i.e. the polarization is orthogonal to the great circle pointing toward the sun). In the following analysis, we focus solely on validating the observed PAs and not the polarization fraction. 

Zenith observations of the twilight sky were taken in the $J$, $H$ and $K1$ bands on 2014 March 24 (see Table~\ref{table:observations}) with the coronagraph optics out of the beam. Orthogonal polarization cubes were made using the standard pipeline recipe, and were then normalized each by its own total intensity to account for the diminishing brightness of the sky over time, before being combined into Stokes data cubes. A mean position angle was then calculated using the mean $Q$ and $U$ values across the field of view for each filter. 

\begin{figure}[ht]
\begin{center}
\includegraphics[width=3.5in]{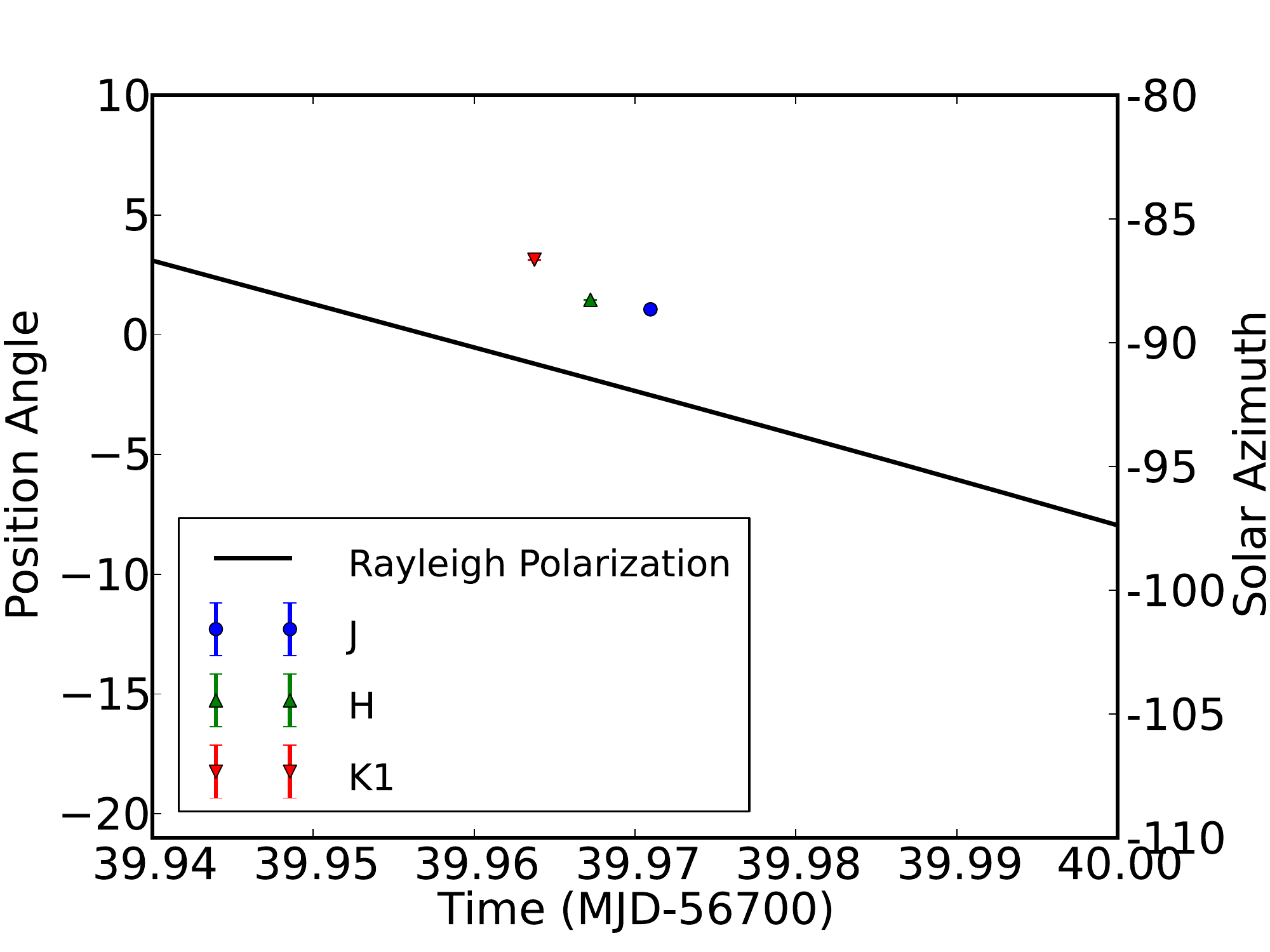} 
\end{center}
\caption{Measurements of the position angle of the twilight sky polarization. The black line indicates the expected position angle assuming pure Rayleigh scattering, based on the solar azimuth as a function of time. GPI measurements in $J$, $H$ and $K1$ differ from the Rayleigh model by $4.34\pm0.02$, $3.30\pm0.01$ and $3.59\pm0.04$ degrees, respectively, where the errors are the standard error of the mean. The formal statistical errors are smaller than the plot symbols in each case. Sunset occurred at $t=$39.949. }
\label{fig:twilight}
\end{figure}

We find the measured PAs are within a few degrees of the expected values as shown in Figure.~\ref{fig:twilight}, confirming there are no sign errors in the pipeline software or instrument problems such as gross angle offsets in the optics. The measurements also reproduce the downward trend over time that is expected from the Sun's motion. However, we find the measured PAs are systematically slightly offset from the expected values by 3.3--4.3\degr. The statistical errors are small when averaged over the 36,000 lenslets; for all three filters the standard error of the mean PA is at or below 0.04\degr, if each lenslet is considered to be an independent measurement. 
The $\sim 4$\degr\ offset may represent an actual systematic bias to measured position angles, but could also be due to some aspect of the sky illumination over Cerro Pachon during these observations, for instance secondary illumination from setting sunlight scattered by clouds on the western horizon. These data were taken between 12 and 30 minutes after sunset when such secondary illumination may be significant. 

Without a larger dataset spanning multiple nights, we cannot be certain yet whether this offset is a recurring systematic. We will continue investigating this issue on future observing runs. 
Twilight observations are somewhat difficult to obtain with GPI; given the small lenslet pixel size and the blue slope of Rayleigh scattering, the polarized near-IR sky becomes undetectably faint very shortly after sundown, making it
impractical to obtain such twilight sky sequences routinely.  Nonetheless these results confirm that GPI can measure PAs to within a few degrees.  We will consider an alternate check of position angles based on circumstellar disk science data itself below in \S \ref{sec:ring_polarimetry}.

\subsection{Instrumental Polarization}
\label{sec:instr_pol}

Observations of polarized and unpolarized stars have allowed us to measure the instrumental polarization induced by GPI's optics \citep{Wiktorowicz2014}. We  describe here one such measurement. The observed instrumental polarization depends on both the telescope itself and GPI's internal optics; since these are fixed with respect to one another there is no need to distinguish between them and so we use ``instrumental polarization'' to refer to their combination. Zemax modeling of the reflections and phase retardances of the optics upstream of the GPI half-wave plate predict an instrumental linear and circular polarization of order $P' \sim 1\%$ and $V' \sim 10\%$, respectively. The dominant effects on incident unpolarized light are expected to be crosstalk into Stokes $U$ and $V$. 

In the instrument reference frame, stellar polarization vectors rotate with parallactic angle. In contrast, the instrumental polarization vector is independent of parallactic angle.  Therefore a sufficiently large range in observed parallactic angle provides a modulation by which stellar and instrumental polarization may be disentangled. This holds true regardless of the level of the stellar polarization, providing a powerful technique by which any sufficiently long science sequence can be inherently self-calibrating for instrumental polarization.
The largest range in parallactic angle obtained during the first night of GPI polarimetry was for the circumstellar disk and exoplanet host star \betapic, for which an $H$-band observing sequence achieved a rotation of $\sim 92^\circ$ in parallactic angle, as indicated in Table 1. In total integrated light, the contribution from the disk is very small. (For instance, \betapic has a measured degree of linear polarization $P = 0.089 \pm 0.024\%$ at position angle $65.2^\circ \pm 8.0^\circ$ east of celestial north in a 700 nm to 900 nm bandpass; Tinbergen 1982). Thus we can use this long $H$-band sequence to enable precise measurement of the instrumental polarization\footnote{The analysis conducted here concentrates on measuring the apparent stellar polarization from the bright portions of the PSF, and is not particularly sensitive to disk-scattered light. The same \betapic data analyzed instead via differential polarimetry will be presented in a future work (Millar-Blanchaer et al., in prep.). }. These observations were taken in AO closed loop, coronagraphically occulted mode with the waveplate rotated between subsequent observations to modulate Stokes $Q$ and $U$ of the beam; the rotation of the pupil with parallactic angle additionally modulated at a different rate the stellar contribution to the beam's polarization.  Each frame was reduced into a data cube containing the two orthogonal linear polarizations split by the Wollaston prism analyzer. Aperture photometry was performed on the occulted PSF in each image, and the fractional polarization $p_{\rm frac} = (I^+ - I^-) / (I^+ + I^-)$  calculated. A model fit to these data as a function of parallactic angle yields both the instrumental and stellar polarizations. 

From these observations, we determine instrumental polarization in $H$ band to be \newline $I \rightarrow (Q', U', P', V') = (-0.037 \pm 0.010\%, +0.4338 \pm 0.0075\%, 0.4354 \pm 0.0075\%, -6.64 \pm 0.56\%)$, where $\pm Q'$ are referenced to the Wollaston prism axes in GPI.  As expected from the Zemax modeling, the dominant terms are crosstalk into $U$ and $V$, and the magnitude of these effects is broadly consistent with expectations. The stellar polarization of \betapic is measured to be $(Q, U, P, \Theta) = (-0.0367 \pm 0.0085\%, -0.0164 \pm 0.0091\%, 0.0402 \pm 0.0087\%, 102.1^\circ \pm 6.6^\circ)$, where $+Q$ is referenced to celestial north. Therefore, our linear polarization measurement of $\beta$ Pic is of the same order of magnitude as the value from Tinbergen (1982) noted above.  After accounting for instrumental polarization, we find the polarization of the unpolarized standard HD 12759 observed with GPI is $(Q, U, P, \Theta) = (-0.016 \pm 0.021\%, +0.036 \pm 0.023\%, 0.039 \pm 0.024\%, 57^\circ \pm 18^\circ)$, which is consistent with zero at 1.6$\sigma$ and has an apparent position angle uncorrelated with \betapic. Thus, instrumental polarization is subtracted well below the $P \sim 0.1\%$ level using the calibration derived from the \betapic data, at least for $H$ band.  For further details and measurements on other standard stars, please see \citet{Wiktorowicz2014}. Additional calibration work is ongoing.

\begin{deluxetable*}{l|llrrr|rrr}
\tabletypesize{\scriptsize}
\tablecolumns{10}
\tablecaption{Observations}
\tablehead{ Target & UT Date & Obsmode\tablenotemark{a} & Int. Time   & \# of. &  \# of & Airmass & Seeing & Field Rot.\tablenotemark{c} \\
                   &      &                          &   (s)  & coadds & exps.\tablenotemark{b} &   & (\arcsec)  &  (\degr) }
\startdata
HD 118666 & 2014-03-24 & J\_coron & 59.6 & 2 & 4 & 1.28-1.29 & 0.7  & 1.3 \\
HD 118666 & 2014-03-24 & H\_coron & 59.6 & 4 & 8 & 1.30-1.32 & 0.7  & 1.3 \\
HD 118666 & 2014-03-24 & K1\_coron & 59.6 & 1 & 4 & 1.26 & 0.7 & 3.1 \\
\hline
Twilight Sky & 2014-03-24 & K1\_direct & 59.6  & 1 & 12 & N/A & N/A & N/A \\
Twilight Sky & 2014-03-24 & J\_direct & 29.1  & 1 & 4 & N/A & N/A & N/A  \\
Twilight Sky & 2014-03-24 & H\_direct & 59.6  & 1 & 4 & N/A & N/A & N/A  \\
\hline
$\beta$ Pic  & 2013-12-12  &  H\_coron   &   5.8 &  10  &  60 &  1.07-1.18 &   0.5     &   91.5 \\
HD 12759     & 2013-12-12  &  H\_direct  &   1.5 &  10  &   8 &  1.05      &   0.45    &    8.7 \\   
\hline
HR 4796A    & 2013-12-12  &  H\_coron   &   29.1 &  2  &  11\tablenotemark{d} &  1.45-1.38 &   0.4     &   2.1 \\
HR 4796A    & 2013-12-12  &  K1\_coron  &   59.6 &  1  &  11\tablenotemark{d} &  1.32-1.27 &   0.4     &   2.2 \\
HR 4796A    & 2014-03-25  &  K1\_coron  &   59.6 &  1  &  12 &  1.01-1.02 &   0.4-0.7 &   33.3 \\
\enddata
\tablenotetext{a}{GPI instrument mode corresponding to choice of filter plus corresponding apodizer, focal plane, and Lyot plane masks. ``direct'' modes have the coronagraphic optics removed from the light path.}
\tablenotetext{b}{Between each exposure the half wave plate was rotated to the next position.}
\tablenotetext{c}{Change in parallactic angle over the course of the observation sequence}
\tablenotetext{d}{One fewer exposure than intended was inadvertently obtained in each sequence, so there are fewer images at 68\degr\ than the other modulator angles, but the data reduction pipeline accomodates this.}
\label{table:observations}
\end{deluxetable*}

\section{Observations of the Circumstellar Disk Around HR 4796A}
\label{hr4796a}

\subsection{Previous Studies}
\label{sec:previous_studies}

\citet{jura91a} first identified HR 4796A, an A0V main-sequence star at 72.8$\pm$1.7 pc, as having significant excess emission at {\it IRAS} wavelengths.  The excess emission was subsequently resolved at  mid-infrared wavelengths into a belt of dust at approximately 50-100 AU radius from the star \citep{koerner98a, jayawardhana98a}.  Near-infrared {\it HST/NICMOS} coronagraphy yielded the first images of the dust in scattered light, revealing a belt or ring inclined to the line of sight by $i$=73\fdg1$\pm$0\fdg2, with projected semi-major axis oriented at PA=26\fdg8$\pm$0\fdg6, and higher surface brightness along the ansae at  1\farcs05$\pm$0\farcs02 radius \citep{schneider99a}.  These observations delivered an angular resolution of 0\farcs12 (8.8 AU) at F110W, and an inner working angle of approximately  0\farcs65 (47.5 AU).   Subsequent ground-based AO images in $H$-band with Subaru/HiCIAO improved the angular resolution and inner working angle to 0\farcs062 (4.5 AU) and 0\farcs4 (29.2 AU), respectively \citep{thalmann11a}. Additional {\it NICMOS} observations at five near-infrared wavelengths (1.7-2.2 $\mu$m) contributed color information indicating that the dust particles are predominantly red scattering \citep{debes08a}. The belt-scattered light has also been detected at optical wavelengths using {\it HST/STIS} with 0.070\arcsec\ angular resolution and inner working angle $\simeq$0.6\arcsec\citep{schneider09a}.  Additional images of the belt in thermal emission have been obtained at $L'$ \citep{lagrange12a} and 10--25 $\mu$m \citep{telesco00a, wahhaj05a, moerchen11a}.  The only polarimetric detection of the belt to date has been a modest detection in $H$-band, where the measured fractional linear polarization of the NE ansa is $>$29\% \citep{Hinkley09a}.

One major area of investigation has been whether or not the HR 4796A belt is geometrically offset from the star in the same manner as Fomalhaut's belt \citep{kalas05a}. The ``pericenter glow'' hypothesis was originally proposed in reference to the tentative finding of greater thermal emission from the NE ansa of HR 4796A as compared to the SW ansa \citep{telesco00a}.  One explanation is that a bound companion (not necessarily a planet and not necessarily interior to the belt) on a non-circular orbit could be responsible for a secular perturbation that creates an eccentric belt with pericenter to the NE \citep{wyatt99a}.  If the belt has an azimuthally symmetric density, then the thermal emission from the NE ansa would be greater due to closer proximity to the star.  However, if the assumption of azimuthal symmetry is dropped, then the belt could be centered on the star and the excess emission could be due to a density enhancement of dust to the NE.  Therefore a direct, astrometric measurement of ring geometry relative to the star was needed to distinguish between these possibilities.  

The {\it HST/STIS} optical (0.2 -- 1.0 $\mu$m) data revealed a 2.9 AU (0.04'') shift along the major axis of the belt such that the brighter NE ansa is in fact closer to the star \citep{schneider09a}.  The Subaru/HiCIAO near-IR observations give offsets of 1.23 and 1.15 AU along the major axis and projected minor axis, respectively \citep{thalmann11a}.  These Subaru observations are best fit by an inclination to the line of sight $i$ = 76.7\degr, which implies the deprojected minor axis offset is 5.0 AU \citep[approximately three times smaller than the stellocentric offset of Fomalhaut's belt;][]{kalas05a}.  These results for the geometric offset of HR 4796A are confirmed in subsequent ground-based studies \citep{lagrange12a,wahhaj14a}, lending empirical support to the pericenter glow or secular perturbation model. 
However, the magnitude of the offset may not be sufficient to explain the difference in grain scattering and thermal emission between the two ansae \citep{wahhaj14a}.

\begin{figure*}[ht!]
\begin{center}
\includegraphics[width=6.5in]{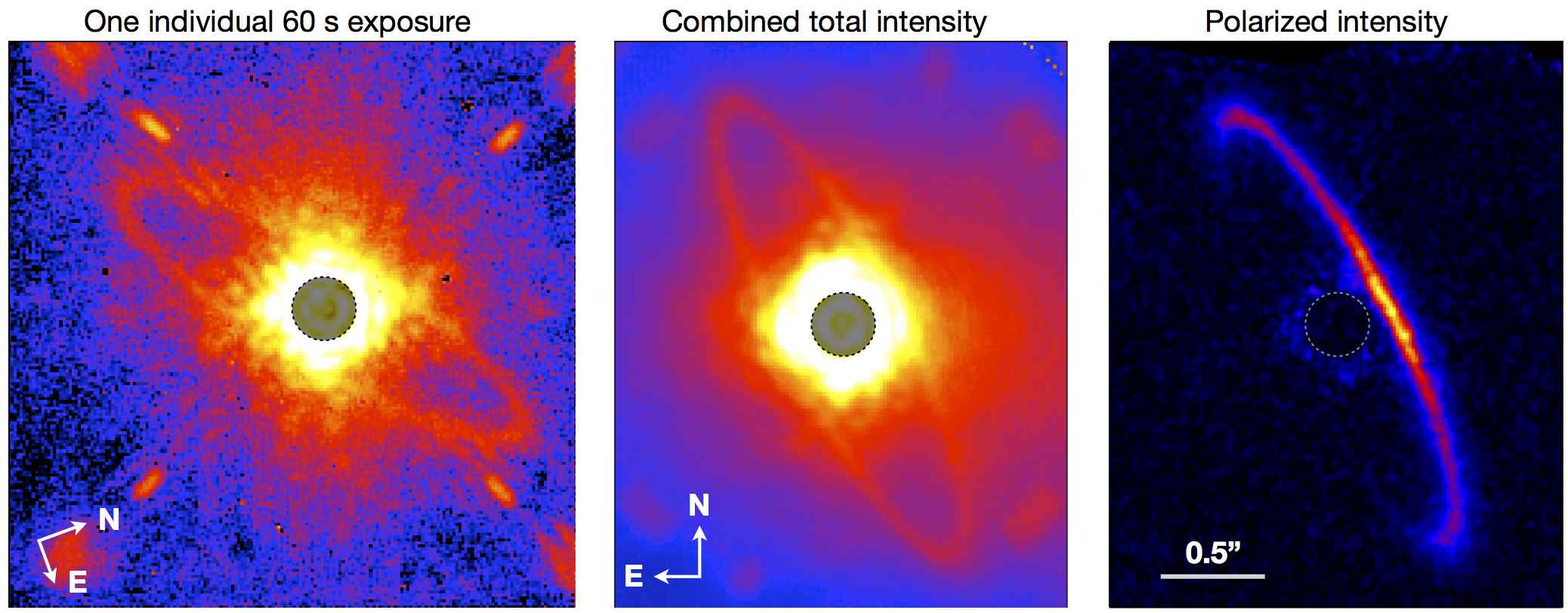} 
\end{center}
\caption{\textit{Left}: A single relatively unprocessed exposure of HR 4796A in $K1$ band for 60 seconds, in one linear polarization, from the March 25th dataset.  The disk is easily seen without any PSF subtraction. Near the corners can be seen the 4 satellite spots used for astrometric and photometric reference to the occulted star, plus four more diffuse spots due to uncorrected ``waffle mode'' wavefront error. \textit{Center:} The total intensity image corresponding to simply derotating and summing the images, as obtained for the 2014 March 25 dataset Stokes $I$ term by inverting the measurement matrix. The excess to the right is dominated by unsubtracted background emission seen only in the $K$ bands. \textit{Right:} The polarized intensity image obtained from the same data. The disk is strongly polarized only on the west side. Note the very uniform and dark background indicating that the stellar PSF has been suppressed to low levels below the noise floor, in this case set by $K$ band background photon noise. Only a very small amount of residual stellar PSF is visible near the center just outside the occulting spot. The image is displayed in the same angular scale in all panels, and the left two have the same brightness display color scale. The overplotted circles indicate the coronagraphic occulting spot size.
\label{fig:raw_and_reduced}}
\end{figure*}

Other key measures of the system include the radial and azimuthal profiles of the belt.  The radial profiles in both thermal emission and scattered light indicate a very sharp belt inner edge and a less steep outer edge.  \citet{lagrange12a} demonstrated that the radial profile of the ring's inner edge along the apparent projected semi-major axis is consistent with a numerical model that assumes an 8 M$_J$ planet with a semi-major axis $a$ = 99 AU that
dynamically clears the interior region of the belt.  Planet detection limits in the infrared have previously excluded the existence of planets located along the major axes greater than $\sim$ 3 M$_J$, although in regions near the minor axes such a planet would not have been detected.  

In addition to the difference in brightness between the two ansa, the ring's surface brightness shows an asymmetry between the east and west sides which has been attributed to a scattering phase function that differs between forward and back scattering.  \citet{schneider09a} model the observed optical brightness asymmetry between the east and west sides of the belt with a Henyey-Greenstein asymmetry parameter $g = 0.16\pm0.06$.  \citet{thalmann11a} find $g = 0.1$ based on their $H$-band data.  It is generally assumed that the small grains that dominate the surface density of the belt preferentially forward-scatter.  Therefore, the brighter east side of the belt has been thought to be the side closer to observers on Earth.  The near-infrared polarimetry presented in this study now calls into question this interpretation.

\begin{figure*}[ht!]
\begin{center}
\includegraphics[width=6.5in]{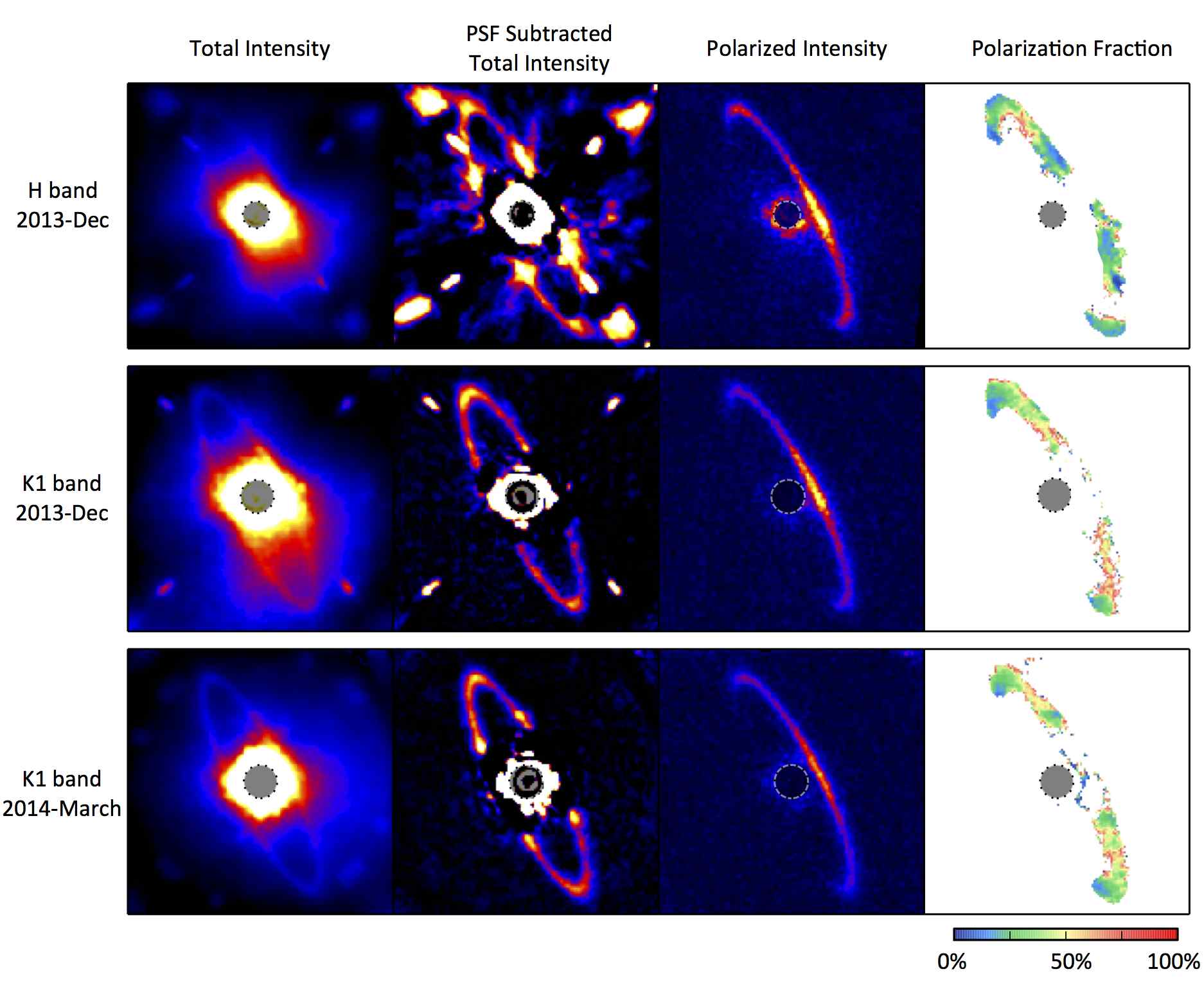} 
\end{center}
\caption{Comparison of the three available datasets including two at $K1$ and one at $H$. The dataset in each row is labeled at left, while the quantites displayed are shown at top. \textit{The left column} shows the total intensity images, without PSF subtraction. Comparing the two $K1$ datasets it can be seen that the March dataset has slightly improved contrast (due in part to observation at lower airmass.) \textit{The second column} shows PSF subtracted total intensity images following the methods in \S \ref{sec:total_intensity_psf_sub}. In the December data with small field rotations, the satellite spots remain as residual systematics in the PSF-subtracted images, but the larger rotation for the March dataset mitigates this and a cleaner PSF-subtracted total intensity image is achieved. \textit{The third column} shows the polarized intensity images. Here despite the differences in wavelength and field rotation the observations are in extremely good agreement. \textit{The fourth column} shows the linear fractional polarization $p=P/I$, with overall general agreement among datasets but limited by the varying PSF subtraction qualities in the second column. Circles again indicate the occulting spot sizes. }
\label{fig:comparison_gallery}
\end{figure*}

\subsection{Observations and Data Reduction for Polarized Intensity}

HR 4796A was observed in polarization mode on the second and third GPI observing runs, in  2013 December and 2014 March respectively, as detailed in Table \ref{table:observations}. 

The 2013 December 12 UT observations occurred just before dawn in good seeing conditions. We obtained brief sequences in both $H$ and $K1$ bands.
The target was acquired via standard GPI acquisition and coronagraph alignment processes \citep{Dunn2014, Savransky2014}. Since the atmospheric dispersion compensator (ADC) was not yet in use, we manually applied position offsets to the calibration unit's low-order wavefront sensor tip-tilt loop to center the star on the occulter at the start of each sequence.  Half way through the $H$ sequence, the IFS cryocoolers were throttled down to reduce their induced vibrations. Modifications to the IFS and AO control laws in early 2014 have mitigated the level of vibration for subsequent observations, and further mitigations are ongoing \citep{Larkin2014, Poyneer2014, Hartung2014}. The waveplate was rotated between position angles of 0, 22, 45, 68\degr\ for the sequence of exposures.   Sky background observations in $K1$ were taken immediately afterwards offset 20\arcsec\ away using the same instrument settings.  
Given the lack of ADC, there is some slight blurring of the PSFs due to atmospheric differential refraction, but the size of this effect is small ($\sim 32$ mas in $H$ band, 15 mas in $K1$ band) compared to the scale of a diffraction limited PSF. 
The disk around HR 4796A is easily visible in individual exposures.

On 2014 March 25 UT, a second set of observations were taken in $K1$ band with the target near transit to obtain increased field rotation.  The polarimetric observations were split in two blocks, with a short series of spectral mode observations sandwiched between them.  The total field rotation across the whole polarimetric dataset is 33\degr.
The observation procedures and total integration time were similar to those in December, including the acquisition of skies immediately afterwards. 
The one significant change is that the ADC was by this time available \citep{Hibon2014}. It was used for these observations as a commissioning test, even though at airmass $\sim 1$ pointing only 10\degr\ from zenith it was not strictly necessary.  As with the December data the disk is easily seen in individual raw exposures.

All observations were reduced using the methods in \S \ref{data_reduction}. Figure \ref{fig:raw_and_reduced} shows the results for the March 2014 dataset, including one individual orthogonal polarization input image, the combined total intensity image resulting from a least-squares average of the input images (i.e., what the differential polarimetry reduction yields for Stokes $I$) and the polarized intensity $P=(Q^2+U^2)^{1/2}$.  In the polarized intensity image, the stellar PSF is highly suppressed and the disk appears as a bright half-ring on its western side. For the first time, the disk can be seen all the way in to its projected closest approach to the star, at least on the western side. In fact the polarized intensity is brightest at the smallest separations. In polarized light the ansae are not particularly brighter than the adjacent portion of the ring to the west, contrary to prior polarimetry at lower contrast and signal-to-noise ratio (SNR) that inferred these were polarized brightness peaks \citep{Hinkley09a}. On the eastern side of the ring, the polarized surface brightness is sufficiently faint that the disk is not detected. 

This polarization asymmetry is a robust result. Looking at the individual exposures in an observing sequence, it is immediately apparent that the disk's surface brightness on the west side is modulated by the wave plate rotation and generally differs between the two orthogonal polarizations; in contrast the east side shows no such modulation.  Furthermore this result is repeatably observed in all three datasets, $H$ and $K1$ in December and $K1$ again in March (see Fig.~\ref{fig:comparison_gallery}). We also note that archival VLT/NACO polarimetry of HR 4796A independently detects this asymmetry (Milli et al., submitted). 
Given the overall consistency of the three GPI datasets, we choose to concentrate on the March $K1$ observations for further analyses since they were observed at lower airmass and with a much greater range of parallactic angle rotation.  Note also that the overall consistency of results between the December and March $K1$ data confirms, as expected, that the ADC (absent in December but present in March) does not substantially change the polarization of transmitted starlight.

\subsection{Polarized Intensity and Position Angle}

From the Stokes $Q$ and $U$ maps, we construct a polarized intensity image $P = (Q^2+U^2)^{1/2}$ that is corrected for the inherent bias resulting from the strictly positive nature of this quantity \citep{Vaillancourt2006}. To reduce the biasing impact of noise, the $Q$ and $U$ images are slightly smoothed by convolution with a Gaussian with FWHM = 2 spaxels (28 mas) prior to computing $P$. Uncertainties in the polarized intensity image, $\sigma_P$, are computed based on the measured background noise in the Stokes $Q$ and $U$ maps, assumed to apply uniformly to all spaxels including those where the ring is located. 

\begin{figure}[ht!]
\centering\includegraphics[width=3.5in]{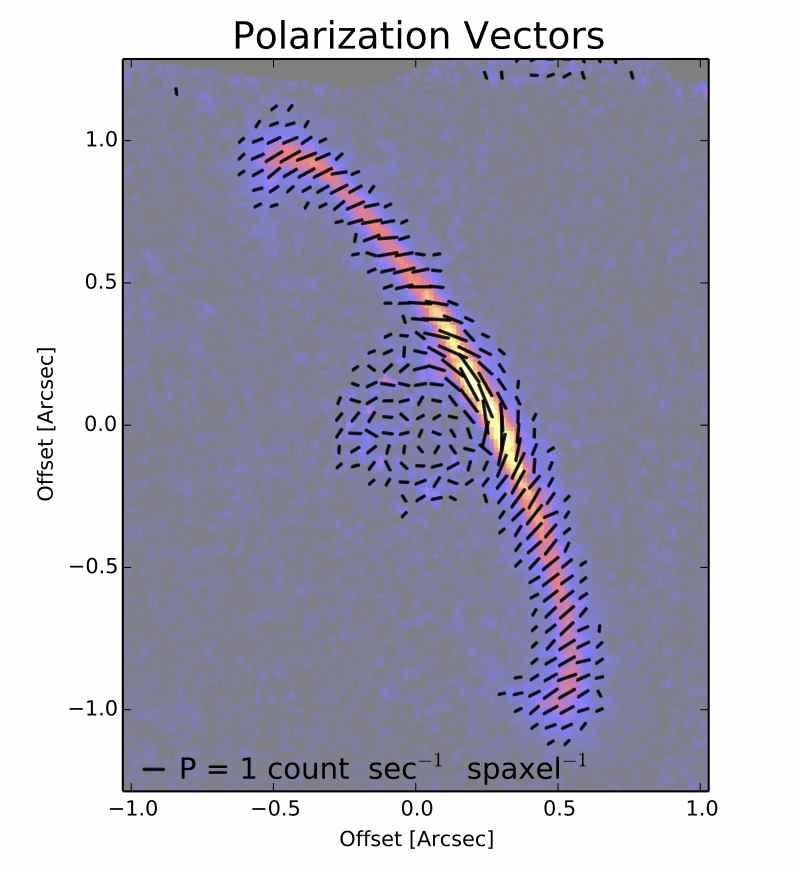} 
\caption{Position angle and intensity of linear polarization around HR 4796A in $K1$ band. The displayed vectors' length is proportional to polarized intensity and orientation shows the position angle for the polarized electric field. Position angles are generally orthogonal to the vector pointing toward the star, as expected. Vectors are shown for polarized intensities $P >  3 \sigma_P$, where in instrumental units $\sigma_P$ is typically 0.1 count sec$^{-1}$ spaxel$^{-1}$ in polarized light. The brightest region near the minor axis reaches up to 2.8$\pm$0.2 counts sec$^{-1}$ spaxel$^{-1}$. Polarization vectors are shown every 4th spaxel. The background image is equivalent to the right panel in Fig.~\ref{fig:raw_and_reduced}, displayed at 50\% transparency.  Qualitatively these position vectors are 
just as expected for polarization of disk-scattered starlight.
}
\label{fig_hr4796_pol_vectors}
\end{figure}

\begin{figure}[htb!]
\centering
\includegraphics[width=3.5in]{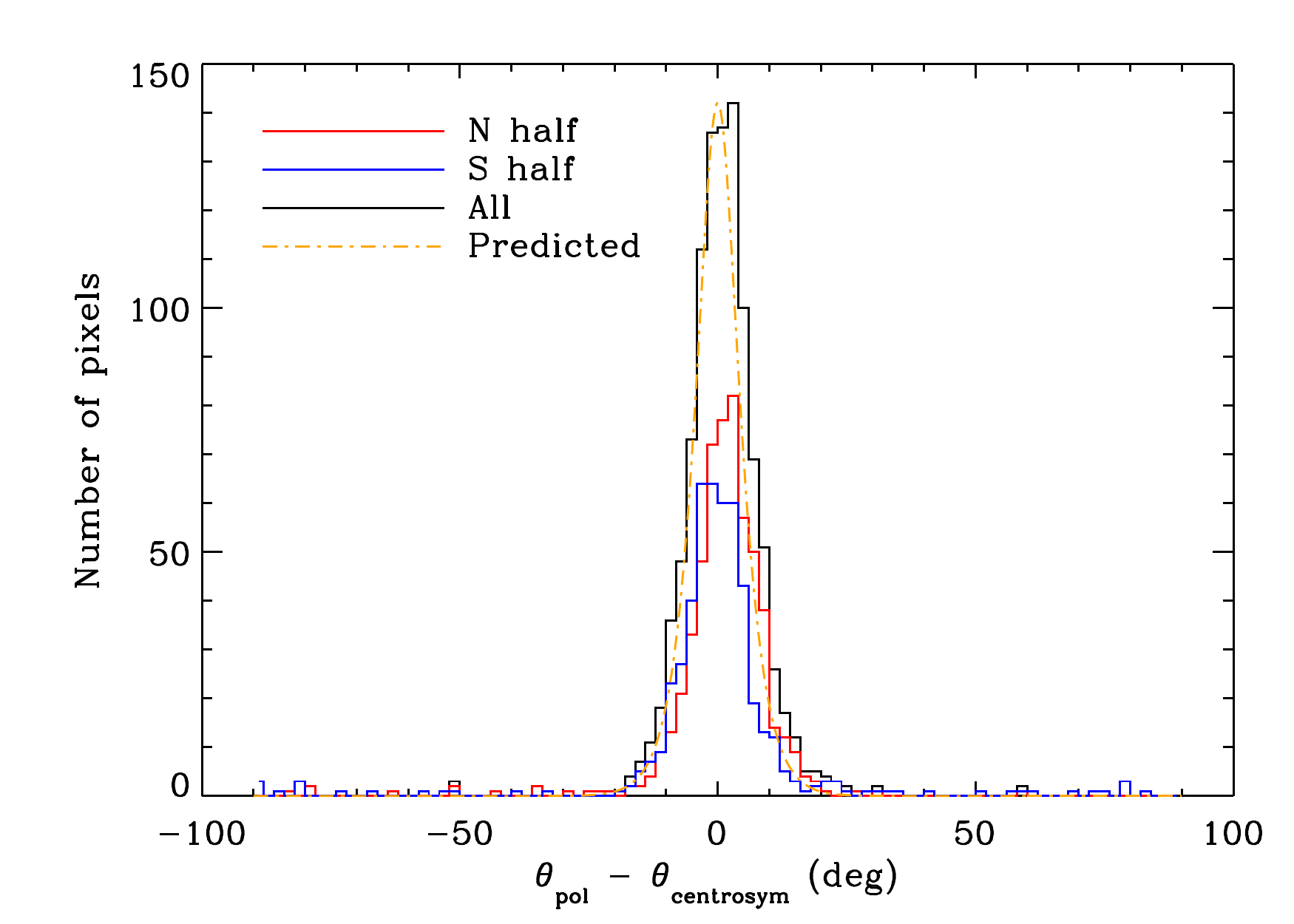} 
\caption{Histogram of the difference in PA between the measured polarization position angles and the predicted values under the assumption that all vectors are centrosymmetric about the stellar position. Only spaxels where $P > 3 \sigma_P$ are included. The black, red and blue solid histograms represent the ensemble of all spaxels satisfying this criterion, those located to the N of the star, and those to the S of the star, respectively. The orange dot-dashed curve is the predicted distribution based on the measured uncertainty on the polarization PA for all spaxels where $P > 3 \sigma_P$. This confirms that the qualitative agreement shown in the prior figure is indeed a precise quantitative agreement as well.
}
\label{fig_hr4796_polar_centrosym}
\end{figure}

From the $Q$ and $U$ maps we also compute the polarization position angle (PA; $\theta = \frac{1}{2} \arctan (U/Q)$) and its associated uncertainty.  Scattering off dust particles results in polarization vectors that are oriented in an either azimuthal or radial pattern.
For all spaxels where $P > 3 \sigma_P$, we find that the polarization vectors are organized in azimuthal fashion, as shown in Fig.~\ref{fig_hr4796_pol_vectors}.  In the March 2014 $K1$ data, the mean deviation of the PAs relative to the expected radial pattern is 0\fdg9 with a dispersion of 5.6\degr, as shown in Fig.~\ref{fig_hr4796_polar_centrosym}. There is a marginal offset of 2\fdg2 in the deviation measured from the northeast and southwest halves of the ring. This may indicate that the location of the star as estimated from the astrometric spots is slightly offset from its true position.
The measured dispersion is only marginally wider than the predicted dispersion of 4\fdg6 based on the distribution of uncertainties on the polarization PA for each spaxel where $P > 3 \sigma_P$.  The position angles are thus consistent with single scattering illuminated by the central star.

Recent astrometric calibration measurements for GPI suggest there may be a $\sim$1\degr\ offset between the coordinate frame of pipeline-processed derotated datacubes and true sky coordinates \citep{Konopacky2014}. Since the Wollaston prism is located after the lenslet array, it must be affected by this offset as well, so the reference frame for Stokes parameters is likely rotated this amount.  For the purposes of checking position angles about the central star, the offset in lenslet position and in position angle $\theta$ will cancel out so the histogram should not be biased. However the above astrometric calibration of GPI is preliminary so conservatively all position angles should be considered uncertain to at least $\pm 1$\degr.

\subsection{PSF Subtraction for Total Intensity}
\label{sec:total_intensity_psf_sub}

\begin{figure*}
\begin{center}
\includegraphics[width=6.5in]{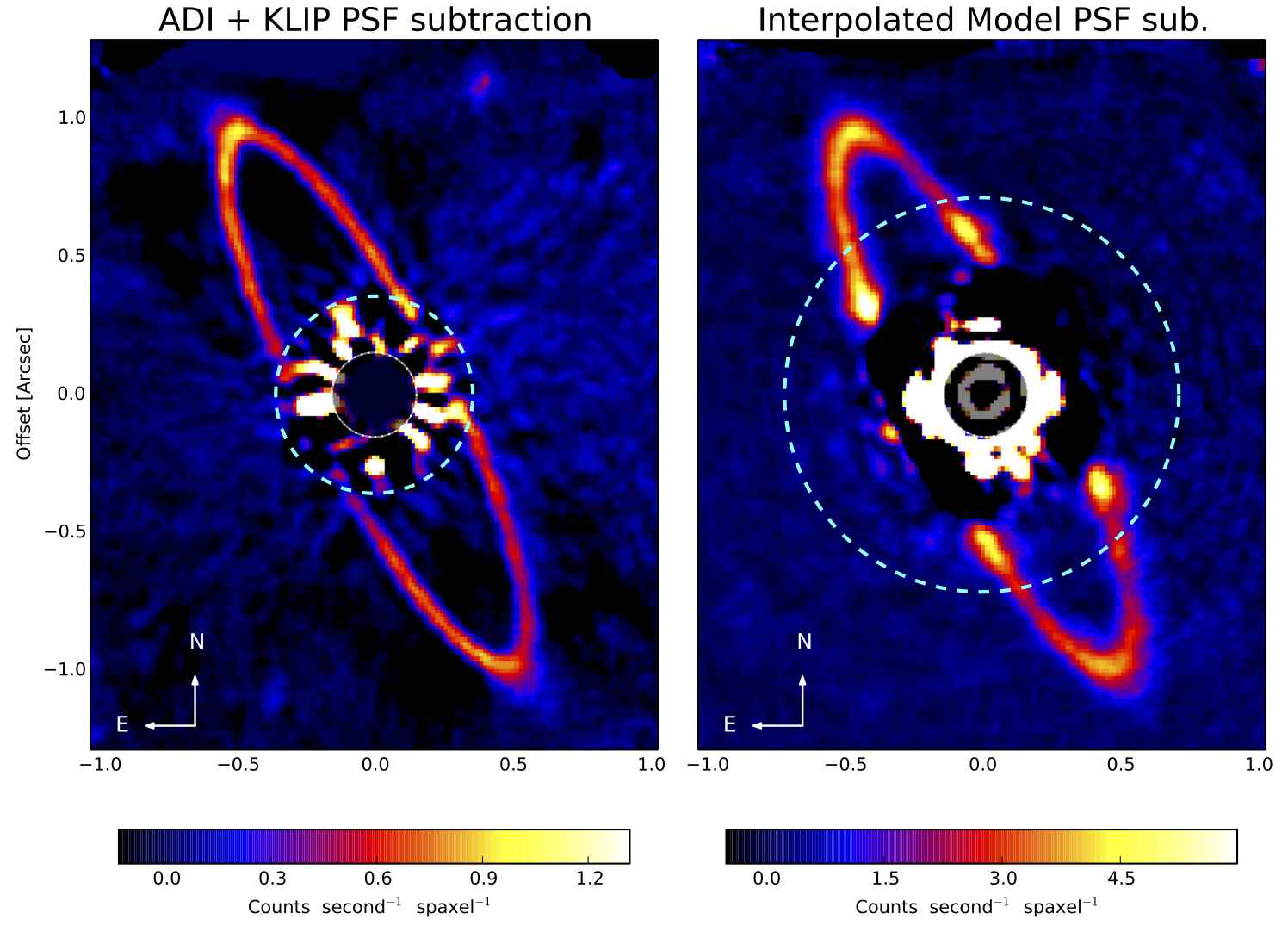} 
\end{center}
\caption{Two versions of PSF-subtracted total intensity images. \textit{Left}: Angular differential imaging (ADI) reduced using the KLIP algorithm, specifically the code implemented as part of the GPI data pipeline. \textit{Right:} An alternate approach using a smooth model for the PSF halo generated after masking out the disk and interpolating across the masked region. See \S \ref{sec:total_intensity_psf_sub} for details. The light blue dashed circled indicate the radii inside of which we consider the data unreliable due to subtraction residuals. The inner dark circle indicates the coronagraph occulting mask, 0\farcs306 in diameter. Note the different display scales as indicated by the color bars. The KLIP-processed image has 2$\times$ better inner working angle but oversubtracts about $\sim$75\% of the flux.}
\label{fig:total_intensity_psf_sub}
\end{figure*}

While the polarized intensity image $P$ is essentially free of residual starlight, the total intensity image $I$ of the ring is affected by significant contamination from the stellar PSF as well as by thermal background in the $K1$ band. In order to measure the polarization fraction, $p = P/I$, we separately reduce the same data via different algorithms to obtain PSF-subtracted images of the disk in total intensity. 

As mentioned in \S3, ADI plus the KLIP algorithm \citep{Soummer2011} is one effective method to remove the PSF. This requires substantial field rotation during the observations, which was only obtained in the March 2014 observations. Reducing these data using the ADI+KLIP implemetation included with the GPI pipeline yields the left hand image shown in Fig.~\ref{fig:total_intensity_psf_sub}. The ring can now be traced in total intensity almost to the projected minor axes. Several noteworthy details can be seen in this image. The ring appears brighter in total intensity on the east side by about a factor of 2; this is particularly visible near the northeast ansa. The brightness does not peak at the ansae as might be expected for an optically thin ring with isotropic scatterers, but rather is brightest in diffuse arcs just to the east of either ansa. Radially outward from either ansa, a more extended surface brightness component is visible. These regions have previously been termed ``streamers'' by some; however they probably do not represent unbound dust particles literally streaming away from the ring, but rather loosely bound particles on eccentric orbits after collisional production in the ring \citep{Strubbe2006}. This smooth component almost certainly extends azimuthally around the entire ring, and is only seen at the ansae due to a combination of the increased optical depth from the line of sight and the spatial filtering action of the PSF subtraction process.

A limitation of the KLIP algorithm is that it is not flux-preserving for either point sources or extended structures such as disks. Due to this systematic bias the apparent suface brightness in the KLIP-processed total intensity image is less than the observed polarized intensity. As a result if we take the ratio $P/I$ we obtain unphysical polarization fractions higher than 100\% for much of the disk.  Forward modeling of the algorithm throughput is one route to mitigating this effect, but in this case since the ring can be easily seen in individual raw images we can take a simpler approach.

In order to obtain a total intensity image in which the ring's scattered light flux is preserved, we developed an independent PSF subtraction method that takes into account the specific geometry of the HR 4796A ring, which only covers a small fraction of the spaxels within the field of view in any given exposure. First, we masked out a region defined by two concentric ellipses that bracket the ring, as well as circular spots at the location of the GPI satellite spots (in the case of the $H$-band data, we masked two sets of four spots, as some of the AO ``waffle mode'' spots at the corners of the dark zone were located close to the ring). We then interpolated the values across the masked regions based on the neighbouring spaxels using a low-order polynomial interpolation scheme (up to 3rd degree in both the $x$ and $y$ directions). Finally, we applied a running median filter to the interpolated PSF to minimize high spatial frequency interpolation artifacts and to avoid stitching issues at the edge of the masked region. The result was an estimated smooth diskless PSF, which was then subtracted from the initial total intensity to produce PSF-subtracted images. This process was applied on a frame-by-frame basis. The resulting images were then rotated to a common orientation, and were median-combined to produce the final PSF-subtracted total intensity image, as shown in the right panel of Fig.~\ref{fig:total_intensity_psf_sub}. Because this procedure does not depend on substantial field rotation, it can be applied to all 3 datasets. The results for the December datasets are shown in Fig.~\ref{fig:comparison_gallery}.

The choice of free parameters for this procedure---polynomial order for interpolation, width of masked region around the ring, and size of the median-filtering window---can significantly alter the resulting total intensity images. The standard deviation of the mean across all images was adopted as an estimate of the uncertainty associated with the total intensity image. We visually inspected the estimated PSF images and standard deviation maps obtained for each observing sequence to identify the optimal parameters and determine the region where the total intensity fluxes seem to be reliably recovered. Based on this inspection, we conservatively estimate that this PSF subtraction method yields good results for the $K1$ observations beyond 50 spaxels (0\farcs7) from the central star. At $H$ band, the proximity of satellite spots and the AO PSF ``waffle mode'' spots significantly reduced the number of usable spaxels in the immediate vicinity of the ring to interpolate the PSF. The lack of significant field rotation during these observations further enhances this issue because systematic errors in the interpolation procedure do not average out as they do in the March 2014 observations. As a result, the final December 2013 $H$ band PSF-subtracted image is of substantially lower quality and we do not consider it further. The two $K1$ datasets yield generally similar results to one another. 

In essence, this interpolated model PSF subtraction has traded larger inner working angle for improved flux conservation. The resulting disk surface brightness is typically 3-5$\times$ higher than in the KLIP-processed images. Computing polarization fractions yields physically plausible values (\S \ref{sec:ring_polarimetry}).  Many of the same features can be seen as in the KLIP image, including the brighter east side, the bright diffuse arcs offset just east from each ansa, and the brightness minimum on the southwest side near PA=$230^\circ$. A decrease in disk surface brightness at this location has previously been reported 
\citep{schneider09a, thalmann11a, lagrange12a}. The ``streamer'' features outside of the ansae are not visible in this reduction, which is to be expected considering that this area was outside of the masked region and thus any disk component there is largely subtracted away.

\subsection{Ring Geometry}
\label{subsec:ring_geometry}

The geometry of the scattered-light ring is expected to trace the locations of dust grain parent bodies and be dependent upon the viewing angles to the system.  The dynamical forces acting on dust particles can also affect the extent of the ring, and gravitational perturbations can induce an offset from the star.
As discussed in \S\ref{sec:previous_studies}, this system has a measurable extent and an offset from the star.  Characterizing these effects can be used to explore the underlying physics affecting the distribution of dust around the star.

We have attempted to constrain the basic geometry of the ring using a simple model fit to the imaging data.
Our model is constructed using an ellipse, with one focus centered on the star, and the corresponding Euler angles describing the orientation of the ellipse.
A small but finite extent to the ring is generated by using an ensemble of ellipses scaled from the reference ellipse.  This scaling is performed so that all ellipses share the same geometrical center.  For simplicity, we assume a flat surface density profile; such a profile has been shown to be inadequate to describe the full extent of the ring \citep[e.g.][]{thalmann11a}, but our primary aim is modelling the ring geometry and we defer more detailed modeling for future work.
We allow for the one-dimensional brightness distribution around the ring to vary, parameterizing the product of the volume density by the scattering phase function with a smooth, periodic spline function of the anomaly.  The computed brightness is scaled by the inverse square law using the distance from the star.
Separate brightness functions are used for the PSF-subtracted total intensity and for the linearly polarized intensity, though the geometrical parameters are shared.
The resulting brightness distribution is smoothed by a azimuthally symmetric Gaussian to represent the unocculted PSF; the characteristic width of this Gaussian is a fit parameter.

The model is jointly fit to the PSF-subtracted total intensity and linearly polarized intensity $K1$-band images from March 2014 (second and third panels in the bottom row of Fig.~\ref{fig:comparison_gallery}, or equivalently the right hand panels of Figs.~\ref{fig:raw_and_reduced} and \ref{fig:total_intensity_psf_sub}).
Both data and model are high-pass filtered prior to comparison, with a Gaussian filter of 20 pixels FWHM.  This mitigates the mismatch of the true unocculted PSF to the assumed Gaussian and filters low-spatial-frequency errors in PSF subtraction in the total intensity image.  For consistency, both total intensity and the linearly polarized image are equally filtered.
Errors are assumed to be uncorrelated and Gaussian in each pixel.  The error level is first estimated by computing a robust estimate of the standard deviation of the image data in 1-pixel annuli.  However, the resulting dispersion is affected by the presence of the disk in the image data.  After a trial fit, the errors are updated using the standard deviation estimate from the residual image data.
In determining the goodness of fit, the model and data are compared over a rectangular region encompassing the disk emission in the high-pass filtered image.  Both the total intensity and polarized intensity images are dominated by residual stellar speckles in the central region, so a star-centered circular region is masked in both images.  The mask radii are 50 and 12 pixels in these images, respectively, identical with the corresponding masked regions shown in Figs.~\ref{fig:total_intensity_psf_sub} and \ref{fig:raw_and_reduced}.
The residuals between the images and models are combined with the error estimates to compute a chi-squared goodness of fit metric.  This is combined with a Levenberg-Marquardt least-squares fitter.  The residuals from the least-square fit show spatial correlation on the scale of image resolution elements. 
We scale the $\chi^2$ metric so that it indicates the error per-resolution-element (rather than per-pixel), and use the resulting best-fit model to seed an ensemble Markov Chain Monte Carlo calculation \citep{foreman-mackey13}.

Using the marginal distributions of samples drawn from the ensemble MCMC calculation, we present the mean parameters and their standard deviations in Table~\ref{table:ring_geo}.  The marginal distributions are Gaussian-like and do not exhibit long tails.
The relatively small error intervals suggest the model is overly constrained; this can arise when there is a mismatch between the model and observations.  In this case, the shape of the assumed PSF and the flat surface density profile may play significant roles.
We have accounted for the astrometric error in scale and orientation using the results of \citet{Konopacky2014}.  Sizes in AU are computed assuming a distance of 72.8 pc, and errors do not account for distance uncertainty.
The offset of the ring is manifested in the nonzero eccentricity of the derived Keplerian orbit.  When combined with the Eulerian viewing parameters, the best-fit offsets of the ring center relative to the star are 0.56 $\pm$ 0.06 AU to the west, and 1.33 $\pm$ 0.13 AU to the south, consistent with previous estimates \citep{schneider09a, thalmann11a}.
The ring is resolved at the ansae by these GPI observations, with a radial extent of $\sim$8 AU (0\farcs11).  However, detailed inferences of the radial brightness distribution are deferred to future analysis.

\begin{center}

\begin{deluxetable}{lr@{ $\pm$ }l}
\tablewidth{2.5in}
\tablecaption{Geometrical fit parameters of the HR 4796A ring.\label{table:ring_geo}}
\tablehead{\colhead{parameter} & \multicolumn{2}{c}{fit value\tablenotemark{a}}}
\startdata
$a_\mathrm{in}$  &    74.43 & 0.61 AU \\
$a_\mathrm{out}$ &    82.45 & 0.41 AU \\
$e$              &    0.020 & 0.002 \\
$I$              & 76\fdg49 & 0\fdg10 \\
$\omega$         & -16\fdg9 & 1\fdg9 \\
$\Omega$         & 25\fdg89 & 0\fdg08 \\
\enddata
\tablenotetext{a}{The uncertainties given here are formal values returned from the MCMC fit. See text for discussion of additional uncertainties.}
\end{deluxetable}
\end{center}

The fitting procedure outlined above provides a first estimate of the ring geometry but makes several assumptions that can affect the reliability of the ring model parameters.
First, the Keplerian description for the shape of the ring is likely incorrect.  Keplerian ellipses will describe the orbits of individual particles, but the ensemble of particles need not follow this description.  The eccentricity controls both the shape of the ring and the offset of the ring center from the star.  Secularly perturbed disks made of ensembles of grain orbits have different behavior \citep{wyatt99a}, so our estimate of the offset may be biased.  A proper physical description will require a revised model.
By using the ellipticity of a keplerian orbit to determine the ring offset, the estimate of inclination can differ from previous studies that relied on the ratio and projected major and minor axes.
We have additionally required the star to be at the focus of our reference ellipse, but an improved description may allow an offset of the center of mass.
The derived parameters rely on astrometric calibration of the position angle and pixel scale.  Additionally, we may be underestimating random errors and neglecting correlations in the data that can also affect the confidence intervals of the fit and derived parameters.
Finally, we have made an assumption of a flat surface density as a function of distance from the star at a given azimuth, while previous works have found steep declines at large separations; however this is partially mitigated by our high-pass filtering of the data prior to model fitting.

\subsection{Polarimetric Scattering Properties vs. Scattering Angles}
\label{sec:ring_polarimetry}

\begin{figure*}[t]
\begin{center}
\includegraphics[width=5.0in]{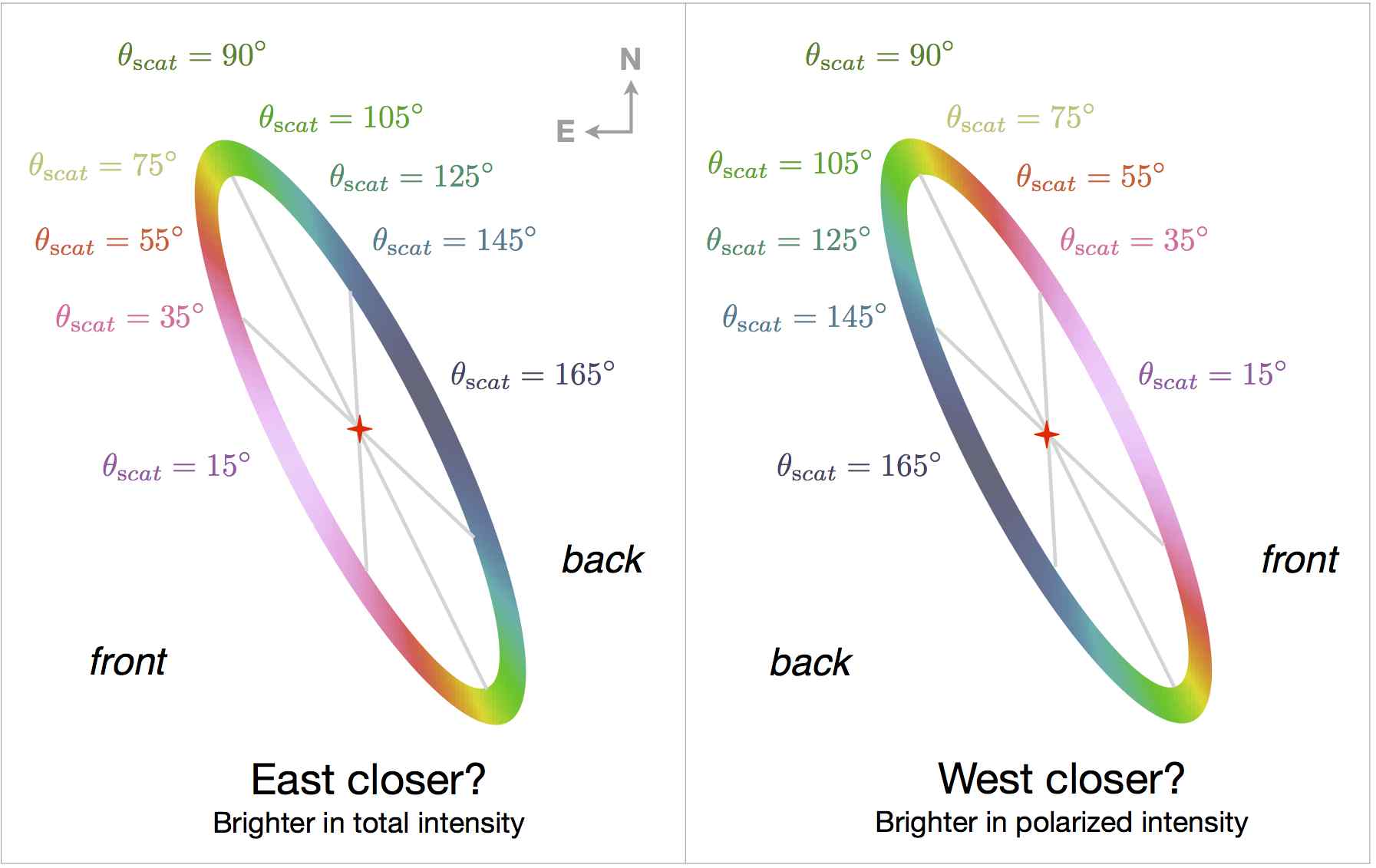} 
\end{center}
\caption{Scattering angles around the ring, for two possible orientations of the HR 4796A disk as observed from Earth. Based on the greater brightness in total intensity to the east and an assumption of predominantly forward scattering, prior works have more often favored the geometry on the left in which the east side of the ring is closer to us. However based on the observed polarized intensities we now argue for the hypothesis at right. The maximum and minimum scattering angles $\theta_{scat}$ are 15\degr\ and 165\degr\ in either case, but flipped between minor axes.  }
\label{fig:scattering_angles}
\end{figure*}

We combine the debiased polarized intensity and interpolated-model-PSF-subtracted total intensity images to generate a map of polarization fraction around the ring. To enable direct comparison with predictions from dust models, we evaluate the scattering angle that characterizes each spaxel in the image based on the location of the star and under the assumptions that the ring is geometrically flat and that the star is coplanar with the ring. Understanding which side of the ring is located closer to us, and thus is characterized by scattering angles $<$90\degr, is necessary to determine the scattering angles. Unfortunately, past observations have not allowed to unambiguously determine this. Generally speaking, the side of a disk that is brightest in total intensity is assumed to be the front side, since scattering off dust particles typically favors forward scattering. In the case of the HR 4796A ring, the east side of the ring is brighter than the west side in the visible and near-infrared, and previous studies have taken that as evidence the east side is closer toward us \citep[][see in particular section 4.4 of the 2009 paper]{schneider99a,schneider09a}. \citet{schneider09a} also found that the ansae brightness peaks are systematically displaced toward the east of the major axis (their Figure 4), which is taken as additional evidence for forward scattering on the east. Our new GPI data reproduce and confirm these features: in total intensity, the disk is brighter on its east side by a factor of up to $\sim 2$, and the brightest peaks of the ansae are east of the  major axis. However, at longer wavelengths the west side becomes equal in brightness to the east side, if not slightly brighter \citep[][Milli et al., submitted]{lagrange12a}, leaving unresolved the ambiguity of which side is closest to the observer. 

In polarized intensity, we see a strikingly different result: the west side is much brighter in linear polarized intensity than the east side.  Indeed, the east side is not detected in polarized intensity over the [35,200]\degr\ range of position angles.  We measure a 3$\sigma$ lower limit of 9 for the flux ratio between the west and east sides near the semiminor axes. The scattering properties of many types of astrophysical dust, across a wide range of assumed compositions and size distributions, have stronger polarized intensities on the forward scattering side, so this result appears to directly contradict the previous hypothesis that the east side is the front. The polarized intensity image yields a much stronger asymmetry than that seen in total intensity. We therefore make the assumption that the west side is closest to the observer and use this convention to determine the scattering angles. If the east side is the closest side instead, the angle we have computed is the so-called ``phase angle'', i.e., the supplementary angle of the actual scattering angle. The reasoning behind this choice is discussed at greater length in \S\ref{sec:analysis}.

\begin{figure}[h!]
\begin{center}
\includegraphics[width=3.5in]{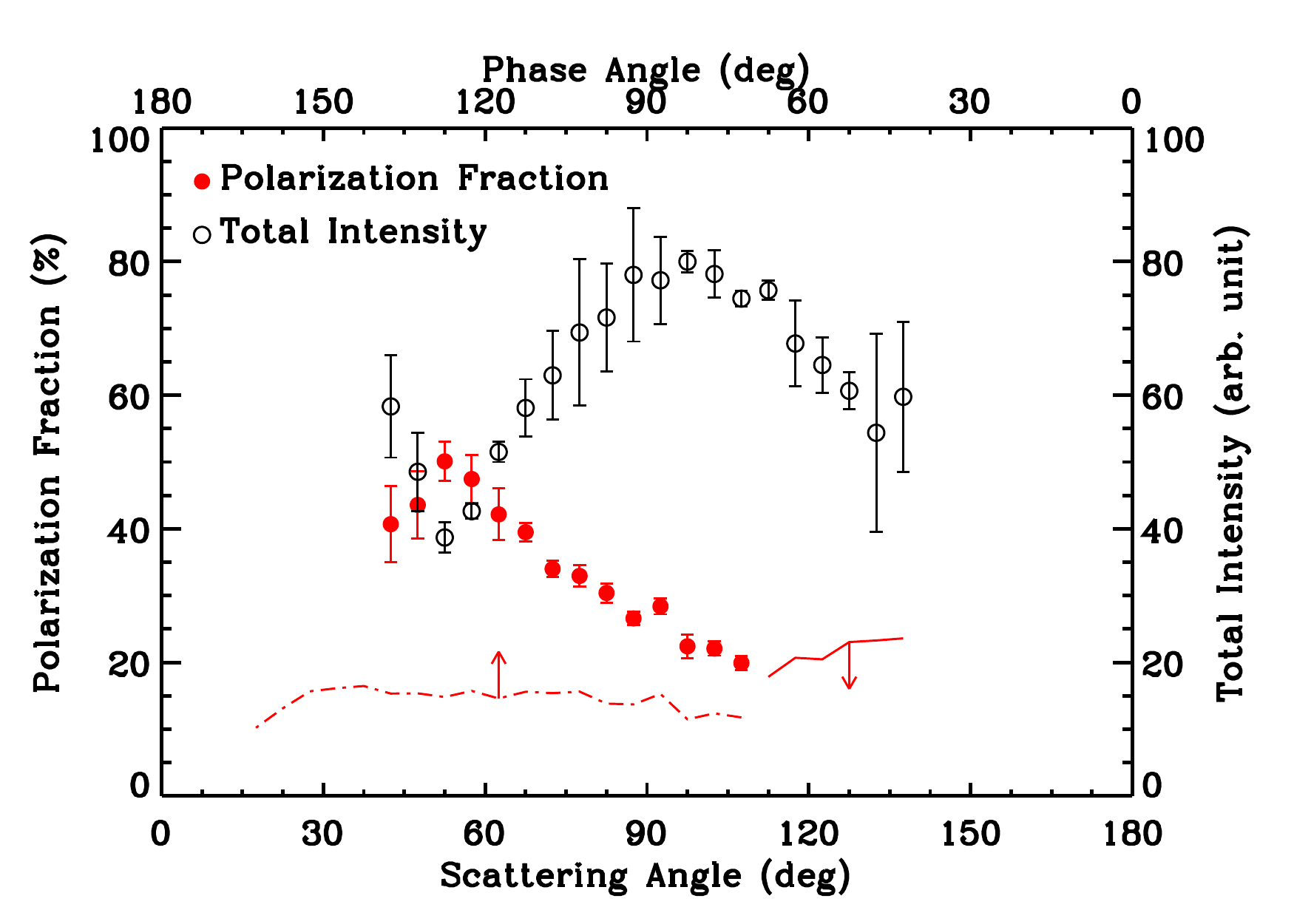} 
\end{center}
\caption{Polarization fraction (red symbols) and total intensity (black symbols) measured along the HR 4796A ring in the March 2014 $K1$-band as a function of scattering angle, assuming that the west side of the ring is the front side from the observer's point of view. The dot-dashed curve represents the 3$\sigma$ lower limit on the polarization fraction estimated by subtracting from the raw total intensity image the minimum flux among all spaxels located at the same stellocentric distance. Over the range of scattering angles where the polarization fraction was estimated, this lower limit is at least a factor of 2 lower than the actual polarization rate, indicating that it is very conservative. On the east side of the ring, too few spaxels have $P > 3 \sigma_P$ for reliable measurements and only an upper limit to the polarization fraction can be estimated, as  represented by the solid curve.}
\label{fig_hr4796_polar_scattang}
\end{figure}

We constructed total intensity and polarization fraction curves as a function of scattering angle in 5\degr\ bins (Fig.~\ref{fig_hr4796_polar_scattang}). We limit the explored range of scattering angles to [40,140]\degr\ to avoid the regions where substantial artifacts may be introduced by the PSF subtraction procedure. Overall, there is relatively little difference between the polarization fractions measured along the NE and SW ansae, as would be expected if the dust properties are uniform across the ring. The largest differences are measured close to our 0\farcs7 exclusion zone radius and may indicate that PSF subtraction is imperfect. Consequently, we compute total intensity and polarization fraction curves that average over the two ansae, and use the largest of the formal statistical uncertainties and the half-difference between the two ansae as the final uncertainty at each scattering angle.

On the east side of the ring for scattering angles larger than 110\degr, too few spaxels have $P > 3 \sigma_P$ and only an upper limit on the polarization fraction can reliably be estimated for this region, as illustrated in Fig.~\ref{fig_hr4796_polar_scattang}. The observed polarization fraction rises smoothly from less than 20\% on the east of the ring to up to 50\% on the west side, with an apparent plateau or turnover for scattering angles $\lesssim$60\degr. 

On the west side of the ring, the polarized intensity can be reliably measured at all scattering angles down to the minimum 15\degr\ allowed by the inclination of the ring. Unfortunately, the total intensity cannot be reliably estimated in the position angle range [230,360]\degr\ due to the strong residual PSF artefacts. Still, a lower limit to the polarization fraction can be estimated from the raw total intensity image. Directly dividing the polarized intensity map by the raw total intensity image yields a strict lower limit of about 3\% along the ring semi-minor axis on the west side of the star. A more useful lower limit can be estimated by taking advantage of the azimuthal near-symmetry of the PSF. Specifically, for each narrow annulus centered on the star, we use the lowest pixel flux as a conservative estimate of the PSF flux at this stellocentric distance and subtract this value from all pixels in the annulus. In essence, the underlying assumption of this method is that the ring does not lie on top of the faintest region of the PSF. If this is true, we then obtain an upper limit on the total ring brightness at each position. From this new ``PSF-subtracted'' image, we can compute a polarization fraction map and use the same averaging in scattering angle bins to produce a lower limit curve for the polarization fraction along the ring. This lower limit is about twice as low as our estimated ring polarization fraction for all scattering angles where we were able to measure it. Furthermore, we find that the minimum polarization fraction along the ring semi-minor axis is about 10\%. The resulting lower limit is illustrated in Fig.\,\ref{fig_hr4796_polar_scattang}.
\clearpage

\section{Analysis of the Ring around HR 4796A}
\label{sec:analysis}

\subsection{Exploratory scattered light models of the HR 4796A ring}

\begin{figure*}[ht]
\begin{center}
\includegraphics[width=\textwidth]{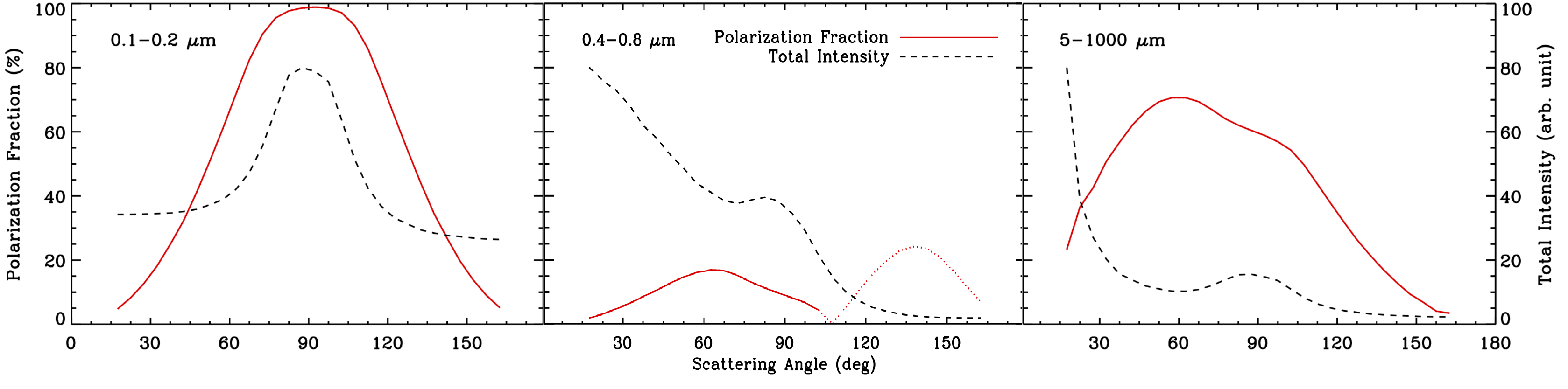}
\end{center}
\caption{Total intensity and polarization fraction curves for three optically thin Mie scattering models using the same astronomical silicate composition but varying the grain size distribution (indicated in each caption legend). These curves can be directly compared with the observed curves for HR 4796A, shown in Fig.\,\ref{fig_hr4796_polar_scattang}.}
\label{fig_hr4796_models}
\end{figure*}

In this section, we explore possible models of the HR 4796A ring that reproduce the observed total intensity and polarization fraction maps derived from the new GPI observations. We limit our discussion to the context of Mie scattering, i.e., to spherical dust grains. We note, however, that elongated but randomly oriented dust grains have scattering properties that approach those of compact spheres for sizes not exceeding about half the observing wavelength \citep{mishchenko92, rouleau96}, suggesting that our results are applicable in for a broader set of assumptions. Synthetic images are computed using MCFOST \citep{Pinte2006}, which provides a full polarization treatment of Mie scattering irrespective of optical depth. In all models, we chose a total dust mass that is extremely low to ensure that the ring is optically thin in all directions, and assume that the density profile of the ring is azimuthally symmetric. For simplicity, we keep the star at the geometric center of the ring, noting that the offset observed in our data is 1) insufficient to introduce intensity asymmetries that exceed our uncertainties, and 2) too small to modify scattering angles by more than $\sim 1\degr$. We also adopt a simple dust composition, namely astronomical silicates from \cite{Draine2003}. Detailed analysis of the SED of the system has allowed \cite{augereau99a} and \cite{Li2003} to characterize in more detail the exact dust composition of ring particles. However, we limit our analysis to the global shape of the intensity and polarization curves, which are relatively insensitive to compositional details. A more thorough analysis is beyond the scope of this first effort and will be presented in a subsequent paper (Fitzgerald et al., in prep.).

For each model image, computed at an inclination of 76\degr, we generated total intensity and polarization fraction curves as a function of scattering angle following the same method as applied to the GPI observations in \S~\ref{sec:ring_polarimetry}. Some representative models, which are discussed below, are shown in Fig.\,\ref{fig_hr4796_models}.

The total intensity phase function derived from the new GPI data is qualitatively consistent with previous studies of the HR 4796A ring at similar or shorter wavelengths, in particular showing the east side to be brighter by a factor of $\sim 1.2-1.5$ and with a brightness enhancement at the ansae. In the context of an optically thin ring, this seems consistent with a model in which the dust particles in the ring are small enough to scatter roughly isotropically in total intensity. For isotropic scatterers, the brightest regions occur at the ansae resulting from the higher line-of-sight optical depth at these locations. Since our observing wavelength is 2~$\mu$m, particles need to be significantly smaller than $\lambda / 2\pi \approx 0.3~\mu$m to scatter isotropically. Such particles are much smaller than the expected blow-out size in the environment surrounding the central A-type star, and it is thus debatable whether a ring model that is dominated by dust that small is physically plausible.
Furthermore, such small particles result in increasingly more isotropic scattering towards longer wavelengths. Therefore, such a model would not predict that the west side becomes brighter beyond 3\,$\mu$m.

As can be seen in Fig.\,\ref{fig_hr4796_models}, a ring model consisting exclusively of small grains (in the 0.1--0.2~$\mu$m range) has a symmetric total intensity phase function with a sharp peak at the ansae, although it seems sharper than is actually observed in the GPI data (FWHM of about 30\degr\ and 60\degr\ for the model and observed curves, respectively). However, such a model produces a polarization curve that is in clear contradiction with observations: for such small grains, the polarization fraction is maximal around 90\degr\ scattering angle and declines symmetrically towards the front and back scattering regimes. Such a symmetric peak is characteristic of scattering by small grains, irrespective of their composition within physically plausible limits. Our observations thus strongly exclude that small grains ($2\pi a/\lambda < 1$, where $a$ is the grain size) are the dominant contributors to scattering in the ring.

Increasing the grain sizes by a factor of a few results in models that simultaneously fail at reproducing the total intensity and polarization fraction curves, as illustrated in Fig.\,\ref{fig_hr4796_models}. As expected for grains with size parameter $2\pi a /\lambda \approx$ a few, forward scattering is heavily preferred and so one side of the ring is several times brighter than the other. This asymmetry is closer to matching the observations in some regards. However, the polarization fraction now presents two peaks, around 60 and 130\degr, with a null around 100\degr. Most notably, the polarization flips sign between the front and back sides for such a model, i.e., the polarization vectors are arranged azimuthaly on the front side but radially on the back side. Therefore, the GPI observations readily exclude that scattering in the HR 4796A ring is dominated by dust particles with size $\lesssim 1\,\mu$m.

Given the wavelength of our observations, increasing the grain size to even larger scales results in dust particles whose scattering properties can be approximated as Fresnel scattering off macroscopic dielectric spheres. In this configuration, a very strong forward scattering peak in total intensity is present but limited to small scattering angles ($\lesssim 30$\degr), which are poorly sampled in our total intensity images at best. Beyond this regime, the phase function has a relatively shallow dependence on scattering angle. In other words, because we sample a limited range of scattering angles, the {\it apparent} phase function could look nearly isotropic even though it is intrinsically very much anisotropic. Furthermore, Fresnel scattering leads to polarization curves which peak at scattering angles much smaller than 90\degr, and as small as 40\degr. Qualitatively, both of these behaviors are consistent with the GPI observations.

To produce a physically plausible explanation for the observations, we construct a model in which grains follow a $N(a) \propto a^{-3.5}$ \citep{Dohnanyi1969} power law distribution and range from 5\,$\mu$m to 1\,mm. The maximum grain size was selected so that even larger grains would not contribute significantly to scattering in the near-infrared or to thermal emission out to 870\,$\mu$m, the longest wavelength at which fluxes are available for this source. On the other hand, the minimum grain size was chosen to match both 1) the approximate blow-out side considering the luminosity of the central star, and 2) the minimum grain size derived from SED analyses of the system \citep[e.g.,][]{Li2003}. It is therefore a physically plausible minimum grain size. We also tried using much larger minimum grain size (50 and 500\,$\mu$m), with no significant consequence on our conclusions, as all grains larger than about 5\,$\mu$m operate in the Fresnel regime limit. 

The resulting total intensity and polarization curves, shown in Fig.\,\ref{fig_hr4796_models}, confirm the qualitative expectations set above. In particular, the polarization is a reasonable match to the observed one, albeit with a 20\% excess at all scattering angles that could indicate that the grains have small-scale surface irregularities. More simply, a grain distribution extending to a somewhat smaller minimum particle size would have a lower peak polarization fraction, in a continuum between the middle and right hand panels of Fig.\,\ref{fig_hr4796_models}. For the 5 $\mu$m minimum grain size, the total intensity curve is relatively symmetric for $\theta_{scat}$ in the range 30\degr--50\degr, and the intensity enhancement at the ansae is broader than in the small grain model. This is more in line with the observed intensity profile. However, the front side is still brighter by a factor $\approx 2$ than the back side in this model, in contradiction with our observations.

\subsection{Interpretation}

This brief exploration of parameter space reveals that the scattering properties of the HR 4796A ring are more complex than may have been expected. While the total intensity phase function is most consistent with small particles, the polarization curve calls for large particles in the Fresnel regime. A bimodal grain distribution cannot explain this behavior since scattering off dust grains cannot contribute polarized intensity without corresponding total intensity. Therefore, this apparent dichotomy suggests that at least one of the hypotheses in our simple models is not fulfilled. Merely modifying the grain size distribution, by changing its slope for instance, is insufficient to account for the contradiction since both observations appear to be characteristic of completely different grain sizes, and yet must instead be driven by the same dust grains. Here we hypothesize on possible explanations for the shortcomings of our exploratory models.

A first interpretation of the apparent dichotomy between the total intensity and polarization curves is that the dust grains contained in the ring do not obey Mie scattering. This would happen if they depart from a compact spherical geometry. Simply increasing the porosity of the grains does not improve the situation, as the polarization curve of porous grains is similar to that of even smaller compact grains. We have computed synthetic observations of a dust model with the same 5--1000\,$\mu$m size distribution but for a 90\% porosity and found that the polarization curve matches closely that of the 0.1--0.2~$\mu$m grain model. Fractal grains, in which the porosity of the grain is organized in a scale-invariant structure, are an expected outcome of low-velocity collisions between small dust grains \citep{Blum2008}. 
Fractal grains, however, have scattered light polarization curves that are bell-like and centered around 90\degr, much like small compact grains \citep{Hadamcik2007}, and therefore do not match the HR 4796A curve.

\begin{figure}[Ht!]
\begin{center}
\includegraphics[width=3.5in]{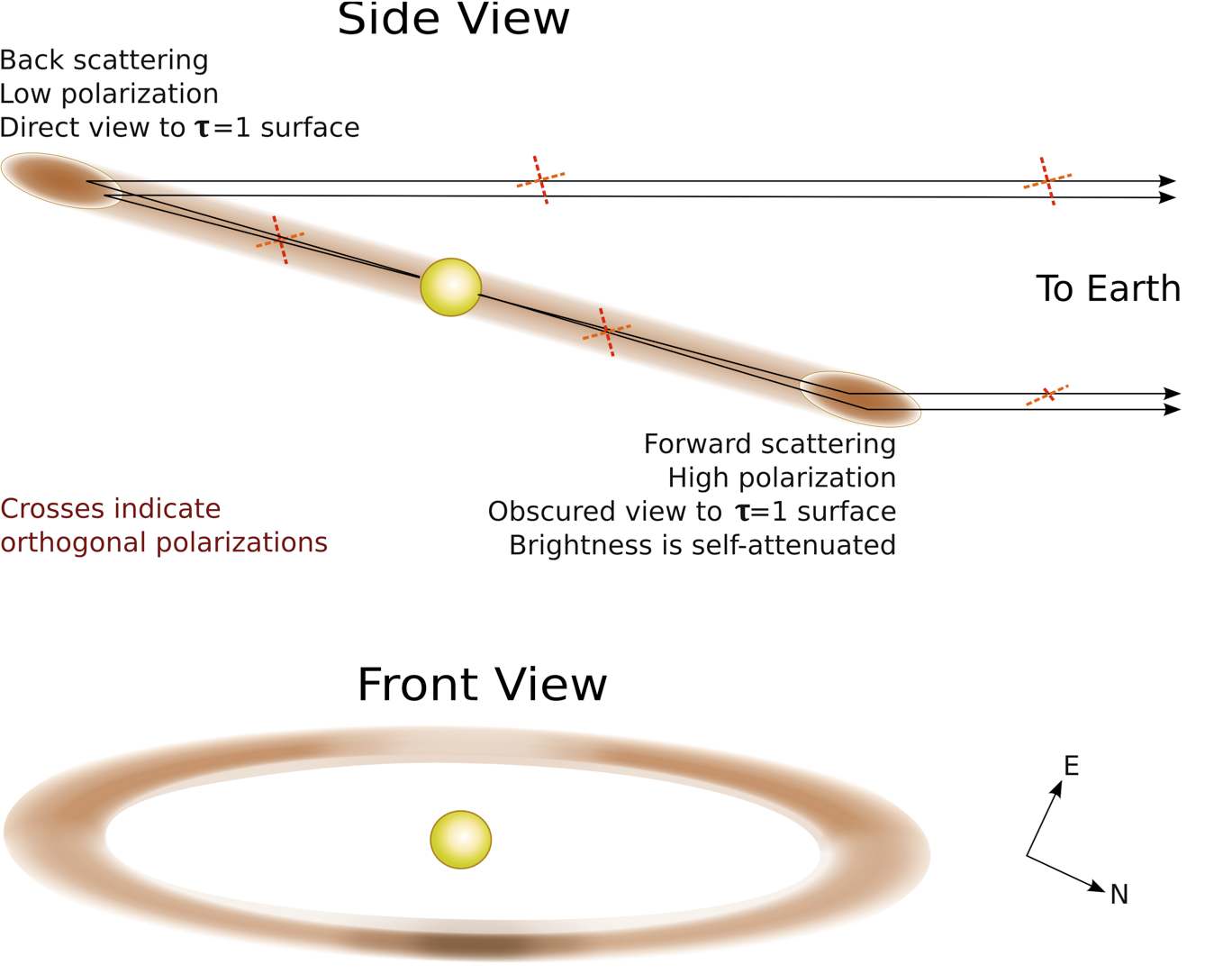} 
\end{center}
\caption{ \label{fig:model_diagram} Cartoon of the geometry and scattering for the ring under the hypothesis that the ring is optically thick. Not to scale, and depicted similar to a reversed color table such that darker indicates higher surface brightness. \textit{Top panel, side view:} Back-scattered light on the far (eastern) side has intrinsically low scattering efficiency and polarization but we have a direct view to the optical depth $\tau=1$ surface where most scattering occurs. Forward scattering has much higher scattering efficiency and polarization; however because the light we observe must pass through the disk, it is attenuated by self-absorption. The portion of the disk's rim for which light can scatter directly to the Earth without self-absorption subtends a small solid angle, also reducing the observed intensity. \textit{Bottom panel, front view:} In total intensity, the less attenuated light path makes the far side appear brighter near the ansae. However closer to the minor axes the intrinsically forward scattering phase function reverses the asymmetry. The increased scattering efficiency at the smallest scattering angles is sufficient to dominate over the self absorption to make the near side minor axis peak in brightness, while the minimal scattering efficiency near back scattering causes the surface brightness to drop near the far side minor axis. In polarized intensity (not shown) the higher polarizing efficiency for forward scattering makes only the near side detectable.
}
\end{figure}

The dust grains in the HR 4796A ring could be elongated. Their scattering properties would depart from those of compact spheres if they are larger than about half the wavelength and/or if they have some degree of correlation in their orientation. For instance, it is plausible that elongated grains are aligned in preferential directions via radiation pressure effects \citep[][and references therein]{Hoang2014}. Considering the much higher complexity of scattering in this situation, we defer a detailed study of such elongated grains to a future paper.

A second possibility is that the ring is not axisymmetric. In the optically thin regime, local density enhancements in the ring readily convert to brightness enhancements, precluding a direct estimation of the scattering phase function from brightness variations along the ring. Our total and polarized intensity images do not reveal clear small-scale asymmetries (clumps or gaps), but smooth variations along the entire ring would be harder to identify. Such global asymmetries could be induced by a planetary mass companion, whose presence is possibly indirectly revealed by the apparent eccentricity of the ring (see \S\,\ref{subsec:ring_geometry}). While total intensity does not track the scattering phase function alone in this situation, the polarization fraction is a robust tracer of the dust scattering properties. The interpretation would thus be that the grains are all large enough to be in the Fresnel regime, as expected from SED analyses. The west side of the ring, where the polarization fraction is highest, would indeed be the front side, whereas the fact that the east is brighter would indicate that its surface density is larger than that of the west side by a factor of $\approx 2.5$, making up for the weaker phase function in the back-scattering regime. Such an extreme density asymmetry may not be physically plausible, and furthermore, this explanation relies on an ad hoc coincendental alignment of the asymmetry with our line of sight. It also does not provide a clear explanation for wavelength dependence of the apparent surface brightness asymmetry.  However, a non-axisymmetric ring cannot be strictly ruled out based on these data. 

A third possible interpretation is that the HR 4796A ring is not optically thin, contrary to general expectations. If the ring is radially optically thick, only the innermost region of the ring is directly illuminated by the star. Because of the high inclination of our line of sight, this region is in direct view of the observer for the back side, but the corresponding region on the front side is hidden from view. Instead the scattered photons received by an observer must pass through an appreciable optical depth of the ring itself and thus are strongly attenuated, resulting in a significantly reduced brightness ratio between the front and back sides of the ring. On the other hand, the polarization fraction curve would not be strongly affected in this context, so long as the optical depth does not exceed $\gtrsim 10$, in which case multiple scattering becomes a dominant factor. To be effective, this scenario only requires a total optical depth of a few along both the radial and vertical directions, which is plausible considering the extremely high luminosity of this debris ring. Total dust masses in the range of 1--100\,$M_\oplus$ have been inferred from SED analyses \citep{augereau99a, Li2003}. If the ring is both radially narrow and vertically thin (on the order of 1\,AU or less), i.e., dynamically cold, such optical depth would be achieved. We defer to a later paper a thorough exploration of this scenario, but consider this is as a plausible explanation for the apparent contradiction between the total intensity and polarization curves observed in the HR 4796A ring.

In fact this hypothesis appears to offer a consistent explanation for several aspects of the ring's observed morphology, as depicted in Fig.~\ref{fig:model_diagram}: 
\begin{enumerate}
\item Consider the ansae. In total intensity the brightest regions are arcs just to the east of the two ansae. Under the above hypothesis, to the west of the ansae the disk is self-shadowing while to the east we can see the directly illuminated inner surface,
explaining both the sign of the asymmetry seen for HR 4796A in total intensity and the offset of the brightest regions away from the ansae.  
\item Furthermore, closer to the minor axes we see different behavior on the east and west sides. Looking at the KLIP-subtracted total intensity in Fig.~\ref{fig:total_intensity_psf_sub} the west side of the ring remains relatively bright as one approaches the minor axis, all the way in to or even just inside the 25-spaxel radius inner working angle. However on the east side, even though the ring starts out brighter near the ansae the surface brightness appears to drop as one gets closer to the minor axis. 
For forward scattering dust we naturally expect the surface brightness to increase as one approaches the forward minor axis and decrease as one approaches the rear minor axis, assuming there are no local nonuniformities in the optical depth through the ring that would mask this. 
Hence the apparent drop in the ring brightness as one approaches the eastern minor axis is consistent with that being the backscattered, intrinsically fainter side.  Note also that since the two minor axes are at essentially the same distance from the star and undergo the same field rotation, it is unlikely that this difference between the east and west side can be ascribed solely to ADI subtraction artifacts. 
\item Lastly this model naturally explains the apparent change in the east/west flux ratio as function of wavelength: Toward longer 
wavelengths $\gtrsim 3~\mu$m the optical depth decreases so the ring becomes less self-shadowing and more light can be scattered through on the west side, while meanwhile the size parameter $2 \pi a/\lambda$ decreases. The net result is that the east side being slightly brighter at $\lambda < 2~\mu$m gives way to more isotropic scattering at longer wavelengths, just as observed.
\end{enumerate}

\subsection{Implications}

If the HR 4796A ring is indeed optically thick, then the intrinsic scattering properties of the particles dominating the optical depth may be well described by Fresnel scattering. This allows us to speculate on how the surface brightness may behave in the regions for which we do not yet have high S/N detections of the disk but which may be revealed by future observations. First, the total intensity phase function for large particles  has an extremely strong peak at the smallest scattering angles; such particles are very strongly forward scattering, with the total intensity increasing by a factor of $\sim 5$ between $\theta_{scat}=40^\circ$ to $\theta_{scat}=20$\degr\ for the 5--1000~$\mu$m case considered earlier. Such a strong peak should be apparent even through a moderate amount of intervening optical absorption such as we hypothesize occurs from an optical depth of a few through the disk.  This implies that at the smallest scattering angles---i.e., near the west side semi-minor axis---the total intensity should increase dramatically.  This area cannot be investigated cleanly using the total intensity data presented here due to the PSF subtraction residuals at these smaller separations. However, we see tantalizing evidence for such a peak in the polarized intensity: the portion of the ring closest to the minor axis, corresponding to scattering angles $\lesssim 20-25\degr$, is unambiguously the brightest location in polarized intensity. This effect is seen even more strongly in the $H$ data than the $K1$ data, as shown in Fig.~\ref{fig:comparison_gallery}.  Symmetry arguments indicate that the polarization fraction must decrease to zero for perfect forward scattering at $\theta_{scat}=0\degr$, so the polarization fraction at 20\degr\ is likely declining compared to larger scattering angles. Indeed, the measured polarization fraction for the ring peaks at $\theta_{scat}=50\degr$ and then decreases at smaller scattering angles in a manner that appears entirely consistent with its reaching zero at 0\degr\ as expected for straight ahead scattering. Linearly interpolating between this boundary condition and the smallest scattering angle for which we have a reliable measurement of the polarization fraction ($\approx$40\% at 40\degr) suggests that the polarization fraction is likely no more than 10--20\% at the disk's western minor axis. For the polarized intensity to peak there, the total intensity must peak even more strongly.  This model predicts that future observations with higher S/N and/or greater field rotation that achieve better inner working angle will find that the total intensity sharply and dramatically peaks at the smallest scattering angles. The western minor axis of the disk likely has the overall maximum surface brightness of the ring.

Secondly, we can apply similar reasoning to the opposite side of the disk at the eastern minor axis where the scattering angles are largest. Our Mie scattering model for a 5--1000\,$\mu$m particle distribution predicts very low brightnesses for both total and polarized intensities, consistent with our nondetections there. If the lower bound of that distribution is instead just a bit lower (as we hypothesized might help also improve the agreement in polarization fraction), then at large scattering angles the polarization fraction may begin to exhibit a so-called ``negative polarization branch'', i.e. a portion of parameter space for which the observed polarization is radial rather than azimuthal in orientation.  Negative polarization branches are a well-known and generally-seen property of dust produced by Solar System comets and asteroids \citep[e.g.,][]{LevasseurRegourd1996,Kiselev1978, Lasue2009,Hines2014}, seen typically for scattering angles\footnote{Note that the Solar System polarimetry literature most commonly uses phase angles $\alpha$ instead of scattering angles $\theta$, so care must be taken to swap complementary angles when necessary. These are related by $\alpha = 180-\theta$, hence scattering angles above 160\degr\ correspond to phase angles below 20\degr.} $\theta_{scat}>160\degr$.  It is interesting to consider whether HR 4796A's ring might exhibit at its eastern minor axis the 90\degr\ reversal of polarization position angles that would be characteristic of a negative polarization branch. The maximum scattering angle of 166\degr\ may be just barely enough to encounter the negative branch, depending upon the exact details of the scatterers. Near infrared observations of some but not all comets show a negative polarization branch \citep[][ and references therein]{Kiselev2014polbook}, reflecting a diversity of dominant particle sizes between different comets. The spectral dependence of observed polarizations is complex but typically the negative polarization branch is more shallow in the NIR than the visible. Thus, even if such a branch is present for the dust around HR 4796A the absolute value of the polarization will be low, just a few percent, which furthermore must be applied relative to the lower total intensity near backscattering.  Detecting such a faint signal may be beyond the reach of even the latest generation of high contrast instrumentation.

In comparison with other circumstellar disks for which measurements of polarized scattering are available over a large range of scattering angles, HR 4796A shows both similarities and some unique aspects. 
Optical imaging polarimetry of AU Mic shows both strong forward scattering ($g\sim0.7$) and high polarizing efficiency that peaks around 90\degr\ scattering (Graham et al. 2007). Based on these data, Graham et al. concluded that the dust must be highly porous to reconcile the high forward scattering and the linear polarizing efficiency. The overall degree of polarization is similar between AU Mic in the visible and HR 4796A at 2 $\mu$m, $\sim 40\%$; both are significantly more highly polarized than can be explained with typical cometary dust.  However, for AU Mic the observations are consistent with Rayleigh-like polarizing efficiency peaking near 90\degr\, which seems strongly ruled out for HR 4796A. Similarly, near-infrared polarization mapping of the AB\,Aur circumstellar disk \citep{perrin09} and the GG\,Tau circumbinary ring \citep{silber00, pinte07} both reveal polarization fractions that peak at scattering angles $\sim100$\degr, i.e., close to the Rayleigh regime 90\degr\ prediction and unambiguously towards the back side of the disks. These few examples confirm that the scattered light polarization properties of the HR 4796A ring are unique among circumstellar disks thus far.

Very few debris disks seen at inclinations similar to that of HR 4796A have been imaged in scattered light to date, leaving little opportunity to compare their respective scattering phase functions over similar ranges of scattering angles. The two closest examples are Fomalhaut \citep{kalas05a} and HD\,15745 \citep{kalas07}, both observed at about 67\degr\ inclination. It is interesting to note that the A-type star Fomalhaut has a belt whose front-to-back scattered light brightness is close to unity whereas the F-type star HD\,15745 has a strongly marked forward scattering preference. In the latter case, \cite{kalas07} concluded that grains sizes in the 1--10\,$\mu$m range were the primary scatterers whereas much larger grains (up to 100\,$\mu$m) have been proposed in the case of Fomalhaut. Such large grains, which operate in the Fresnel regime are consistent with our conclusion regarding the HR 4796A ring. Unfortunately, neither of these two other disks has been detected in polarized scattered light, precluding a more complete comparison with HR 4796A.

Given the known total dust mass inferred from longer wavelength observations, for the HR 4796A ring to be optically thick as we hypothesize, it must be geometrically thin and dynamically cold. Dynamical models are beyond the scope of this current study.\footnote{Note added after ApJ submission: It has recently been brought to our attention that \citet{2008A&A...481..713T} have previously suggested that a dynamically cold model for HR 4796A provides a good fit to the sharp outer edge of the ring.} Rings are a relatively common form for debris disks, and based on the similarity of orbital radii and characteristic temperatures, the Kuiper Belt is often taken as a basis for comparison. 
However, the Kuiper belt is strongly dynamically stirred by its interaction with Neptune, and a substantial fraction of its constituent planetesimals are on highly inclined and/or eccentric orbits. 
Fully a third of currently known trans-Neptunian objects have orbital inclinations greater than 10\degr.
A better Solar System analogy might be one of the rings of Saturn, which are optically thick while having a very small scale height.
The narrowest rings are shepherded by the gravitational influence of small moonlets orbiting just inside and outside of the ring. It is tempting to speculate that a similar process is at play around HR 4796A. Previous work has found that the sharp inner and outer edges of the disk are potentially explainable using the dynamical influence of two planets, one relatively close inside the ring and the other outside \citep[e.g.,][]{lagrange12a}. Dynamical modeling of these new observations from GPI will help test this scenario.

\section{Summary and Future Prospects}

The first light observations of HR 4796A presented here demonstrate the discovery potential of GPI.
The high performance AO system and coronagraph enable the ring to be seen clearly in total intensity images both without and with PSF subtraction. Differential polarimetry not only provides dramatic starlight suppression by two orders of magnitude or better, it also is yielding new astrophysical insights into the constituents of circumstellar disks through constraints built on measurements of the scattering phase functions. 

The first-of-its-kind integral field polarimetry architecture of GPI is optimized for high contrast polarimetry. While this architecture -- and the fact that GPI always observes in ADI mode -- does complicate some aspects of data reduction, we have developed the necessary methods and shared them with the community as part of the GPI data reduction pipeline. The calibration observations presented here for starlight suppression and instrumental polarization are part of a long term effort to comprehensively characterize GPI's performance as a polarimeter.  These early results establish clearly that GPI is performing well in terms of polarimetric accuracy and precision. 

Having a well-understood instrument is important when considering the HR 4796A observations, because the results are at odds with expectations drawn from previous studies. The side of the disk that is fainter in total intensity is $\gtrsim 9$ times brighter in polarized intensity. The relatively isotropic total intensity phase function suggests scattering by small particles in the Rayleigh regime, but this is strongly ruled out by our observations that show the polarized intensity does not peak at the ansae but instead at the western minor axis. We have not been able to identify any simple model that can reconcile these observations. Instead, we must consider more complex scenarios. Our favored model is that the disk is optically thick, and absorption or self-shadowing on the near side reduces its apparent brightness, making the intrinsically brighter forward scattering on the west masquerade as fainter back scattering.  
In this optically thick, geometrically thin scenario, only at the smallest scattering angles does the phase function's strong forward scattering peak come to dominate over the self absorption. This effect is only hinted at in the present data, however.

Our follow-up study (Fitzgerald et al., in prep) will present deeper multi-wavelength polarimetric data of the HR 4796A system that have a greater capacity to measure the total intensity of light around a larger range of angles, improving measurement of the dust scattering properties.  The next steps in modeling the system  will better constrain the geometry and consider a wider variety of potential scatterers.  Additional imaging over a larger range of wavelengths with complementary instruments (e.g. SPHERE/ZIMPOL, MagAO) will also help test these models. HR 4796A has lately become somewhat of a poster child for new high contrast AO systems, and for improved reprocessings of archival datasets. This not only provides a fair test case for comparing instrument performances on spatially extended targets (akin to the diverse collection of images of HR 8799's planetary system observed by essentially every AO-equipped large telescope), it also means that there will be an extraordinarily rich multiwavelength dataset to interpret.  Vega famously was a calibration star that turned out to be an unanticipated circumstellar disk host; HR 4796A appears to be an extreme AO system test target which has turned out to be more astrophysically complex than was previously expected.

The data presented in this paper were obtained during the first two observing runs in which the polarimetry mode of GPI was used. Since then, polarimetry has been extensively exercised in the ongoing commissioning program, and also in the shared-risk early science observing run in April. Approximately half the observing programs during that run were polarimetry, studying not just circumstellar disks but also mass loss around massive post-main-sequence stars and accretion and outflow around FU Orionis stars. GPI will also provide a tool for spatially resolved polarimetry of solar system bodies such as asteroids and Saturn's moon Titan. 
In late 2014 the GPI Exoplanet Survey key science program will commence a comprehensive survey of 600 nearby stars looking for Jovian exoplanets; a search for debris disks around a subset of these stars with known infrared excesses is a key part of the campaign. 

The polarimetry mode of GPI is complementary to its planet detection capabilities. Through imaging spectroscopy, GPI will detect and characterize the largest bodies in these planetary systems, which gravitationally dominate the dynamics of the regions outside a few AU. Conversely the polarimetry mode of GPI will provide sensitivity to the much smaller planetesimals and minor planets that populate debris disks and rings, and give rise to observable dust through collisional grinding. Correlations between the presence of planets and disks are increasingly seen; observations with GPI, SPHERE, and other high contrast AO systems will probe these correlations in greater detail and test theories for planet-disk interactions. The proposed WFIRST/AFTA Coronagraph offers one path to extending these capabilities several more orders of magnitude in contrast, to approach the point of being able to detect Solar-System-like dust levels with $\tau$ much less than that of the relatively massive and dusty HR 4796A ring. A polarimetric capability as part of the AFTA Coronagraph would be an unprecedented platform for studies of exozodiacal light arising from asteroid belts and comets around nearby stars, which would in turn inform on the astrophysical backgrounds for future terrestrial planet imaging missions.  Together the sensitivity of these instruments to massive Jovian planets and ensembles of micron-sized dust particles will enable the characterization not just of individual planets, but of planetary systems overall.

\acknowledgements

M.D.P. and J.R.G. led the development of GPI's polarimetry mode including design, implementation, and data analysis methods, and M.D.P. implemented the polarimetry reduction pipeline and led the first light observations with GPI polarimetry. G.D. led the analysis and interpretation of the HR 4796A scattering properties in collaboration with M.D.P., M.P.F., and others on the team. M.M.B. performed instrument calibrations, contributed substantially to reduction software, reduced data, and conducted the contrast and sky polarization analyses. M.P.F. performed the analysis of the ring geometry. S.J.W. analyzed instrumental polarization. P.K. and T.M. summarized previous observations of HR 4796A. B.M. provides overall scientific leadership to the GPI team. The remaining authors collaborated in the development and commissioning of GPI.  R.D.R and J.P. provided a careful reading and internal review of this paper prior to submission.

We thank the international team of engineers and scientists who worked to make GPI a reality.  
The Gemini Observatory is operated by the  Association of Universities for Research in Astronomy, Inc., under a cooperative agreement with the NSF on behalf of the Gemini partnership: the National Science Foundation (United States), the National Research Council (Canada), CONICYT (Chile), the Australian Research Council (Australia), Minist\'{e}rio da Ci\^{e}ncia, Tecnologia e Inova\c{c}\~{a}o (Brazil) and Ministerio de Ciencia, Tecnolog\'{i}a e Innovaci\'{o}n Productiva (Argentina).
We acknowledge financial support from the Gemini Observatory, the National Science Foundation (NSF) Center for Adaptive Optics at University of California, Santa Cruz, the NSF (AST-0909188; AST-1211562), NASA (NNX11AD21G and NNX10AH31G), the University of California Office of the President (LFRP-118057), and the Dunlap Institute, University of Toronto. 
Portions of this work were performed under the auspices of the U.S. Department of Energy
by Lawrence Livermore National Laboratory under Contract DE-AC52-07NA27344, and other portions under contract with the California Institute of Technology Jet Propulsion Laboratory funded by NASA through the Sagan Fellowship
Program executed by the NASA Exoplanet Science Institute.  We also acknowledge support from the Natural Science and Engineering Council of Canada. MDP was supported in part by a National Science Foundation Astronomy and Astrophysics Postdoctoral Fellowship, NSF AST-0702933. MDP also acknowledges support from the STScI Director's Discretionary Research Fund.

\appendix

\section{Polarimetry hardware details for GPI}

\label{appendix:details}

\textit{Beamsplitter:}
As noted in \S \ref{section:ifp}, the optimal design for a Wollaston prism for integral field polarimetry has the spots separated by 45$\degr$ relative to the lenslet grid and interlaced like a chess board. The desired separation of the spots should be $l/\sqrt{2}$ where $l$ is the projected separation between lenslet centers, yielding for GPI a separation of 7.1 pixels = 126 $\mu m$. 
Taking into account the 235 mm focal length of the IFS camera optics, a divergence angle of only 0.03\degr\ between polarizations is required.  This is achieved using an MgF$_2$ Wollaston prism with a 1.4\degr\ prism angle. 
GPI's lenslet array is rotated by 26.5\degr\ relative to the detector to avoid overlap of spectra in spectral mode, and therefore the polarization dispersion direction must be -18.5\degr\ or 71.5\degr\ relative to the detector. The Wollaston is therefore mounted at an angle of -18.5\degr\ on the prism exchange translation stage; this is worth emphasizing because it sets the native reference frame of GPI polarimetry which is neither aligned with the IFS lenslet array rows and columns, nor with the orientation of the AO dark hole, nor with any particular sky reference frame. That translation stage uses magnetic latches at opposite ends of its travel to ensure precise position repeatability for the spectral and Wollaston prisms. See \citet{Larkin2014} for further details on the GPI IFS and its mechanisms.

\textit{Modulator:}
An achromatic MgF$_2$/quartz wave plate was procured from Bernhard Halle with $\lambda/2 \pm 4\%$ retardance from 0.7-2.5 $\mu m$.   To achieve accurate polarimetry with GPI it is not required that the retardance of the waveplate be precisely 0.50 waves at all wavelengths, but rather that its retardance vs. wavelength is accurately known. Details of the wavelength dependent retardance may be found in \citet{Wiktorowicz2014}. This waveplate was mounted in a custom rotation stage physically located in GPI's calibration interferometer subsystem, in the collimated beam shortly before the IFS entrance window. This rotation mechanism achieves angular position repeatability of 0.023$\degr$ (3 $\sigma$) in laboratory testing at room temperature.  Temperature dependent bias in the mechanism's home position is tested to be $<0.25\degr$ between -12.5 \degr C and +25 \degr C. 

The wavefront error of the transmitted beam through the modulator is 47 nm peak-to-valley and 11 nm rms. For comparison the entire GPI optical path to the coronagraph focal plane was measured in the lab to acheive 13 nm rms, with individual optics $< 1$ nm rms. The substantial wavefront error of the waveplate in this context supports our decision to place it after the coronagraph focal plane.

\textit{Bandpass Filters:}  GPI contains five spectral bandpass filters, the details of which were motivated by the needs of exoplanet spectroscopy. Three filters correspond to the $Y$, $J$, and $H$ bands, though the specific wavelength ranges do not correspond precisely to standard filter bandpasses such as the MKO filter sets. Because the spectral dispersion of the GPI IFS increase toward longer wavelengths, it was necessary to split the $K$-band into two overlapping filters to avoid spectral overlap on the detector. Thus for polarimetry beyond 2 $\mu$m we have a choice between two custom filters, $K1$ from 1.9--2.19~$\mu$m  and $K2$ from 2.12--2.38 $\mu$m. Additional documentation on the spectral response of these filters are available on the instrument team's web site\footnote{http://docs.planetimager.org/pipeline/ifs/filters.html}, and for the overall spectral response of GPI see \cite{Maire2014}.

\begin{figure}[ht]
\begin{center}
\includegraphics[width=6.5in]{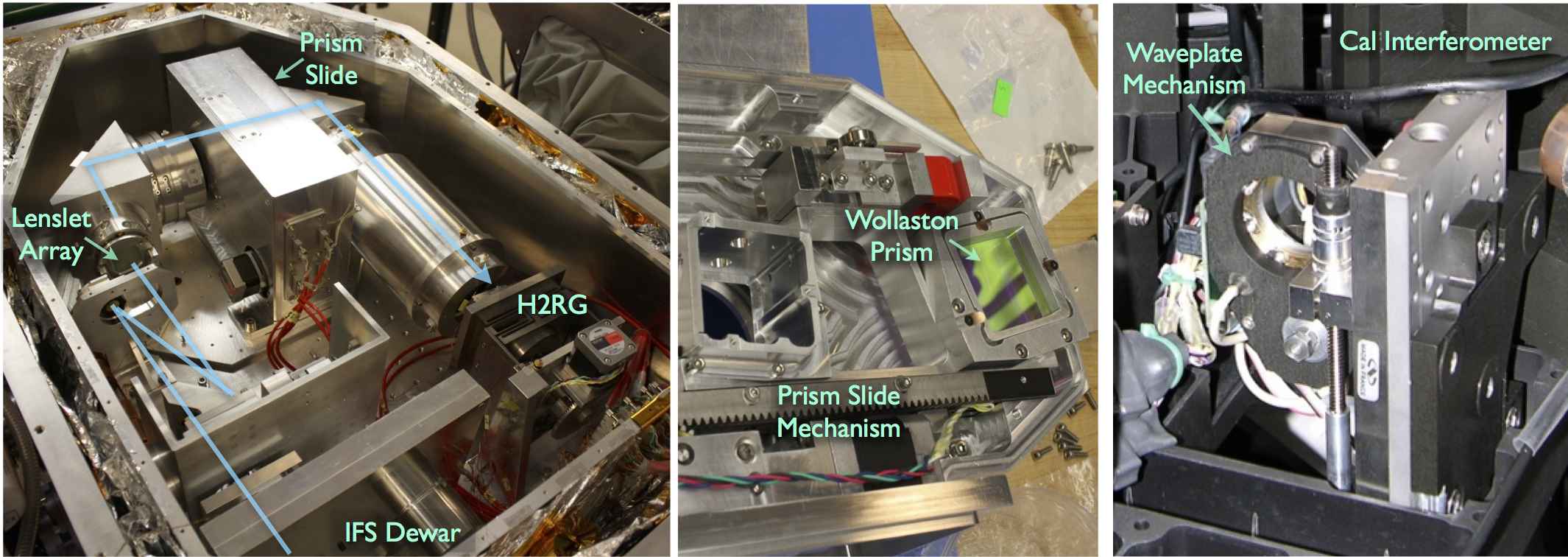} 
\end{center}
\caption{Assembled GPI hardware highlighting several key components of its polarimetry mode.
    \textit{Left:} The IFS dewar showing the internal light path, in blue, and mechanisms.  The prism slide is located inside the rectangular enclosure along with the filter wheel. The dispersed light is focused onto the Hawaii 2RG detector at lower right. \textit{Center:} A view of the prism slide enclosure with the cover opened, showing the Wollaston prism mounted indicated on the right. The spectral prism has since been mounted in the other prism holder at left. \textit{Right:} The waveplate insertion and rotation mechanism is the hexagonal structure shown here mounted into the calibration interferometer enclosure. 
        }
\label{hardware}
\end{figure}

\section{Data Analysis Methods for Integral Field Polarimetry}

\subsection{A Least Squares Approach to Sequence Combination}

\label{appendix:leastsquares}

Astronomical polarimeters often adopt a data reduction method in which successive images are taken with polarization states modulated to swap e.g. the $+Q$ and $-Q$ polarizations between the two orthogonal polarization channels. A double-difference image constructed from these will be relatively robust against both atmospheric speckles and instrumental polarization effects occurring after the modulator. 
That traditional double-differencing reduction will not suffice for GPI with its constantly rotating orientation with respect to the sky:  it is not possible in general to choose a single set of waveplate rotation angles such that sums and differences of the input images would correspond to Stokes parameters in a fixed celestial frame. While specific waveplate rotation angles to achieve any desired modulation could be calculated as a function of time and telescope orientation for each target, that approach would introduce unnecessary complexity and timing dependency into GPI observations.

Instead, the traditional double differencing method can be considered a special case of the more general approach in which the astronomical polarization signature is recovered by constructing the equations of condition for $N$ measurements at arbitrary modulator positions $\theta_1$ to $\theta_N$: 

\begin{equation} 
\left( 
\begin{array}{c}
I'_1 \\
I'_2\\
\vdots \\
I'_N\end{array} 
\right)
= 
\left( 
\begin{array}{cccc}
m_{11} (\theta_1)  & m_{12} (\theta_1)  & m_{13} (\theta_1)  & m_{14} (\theta_1)   \\
m_{11} (\theta_2)  & m_{12} (\theta_2)  & m_{13} (\theta_2)  & m_{14} (\theta_2)   \\
\vdots &  \vdots & \vdots & \vdots \\
m_{11} (\theta_N)  & m_{12} (\theta_N)  & m_{13} (\theta_N)  & m_{14} (\theta_N)  \end{array} 
\right)
\left(
\begin{array}{c}
I \\
Q \\
U \\
V\end{array} 
\right)
 = M_{meas} \cdot \bf{S}.
\label{Eq-least-squares}
\end{equation}

If the matrix elements $m_{ij}(\theta)$ are sufficiently well known, the input polarization vector may be solved for via the method of least squares.  The double-differencing method simplifies this reduction to convenient addition and subtraction by enforcing a particular selection of the modulation angles $\theta$, but the general approach holds even in cases where that simplication is not available. 

Using the known instrumental response matrix, for every exposure in a sequence we write down the Stokes vector equation describing that measurement. We then solve this complete set of equations of condition to derive the astronomical polarization for each point in the field of view.  We obtain the terms of the measurement matrix $M_{meas}$ by modeling the instrument's performance using the standard Mueller matrix approach.  Each component in the system is described by some Mueller matrix, and the overall $M_{GPI}$ is the product of these individual terms:
\[
    M_{GPI}(\theta,\phi,\eta) = M_{Woll\pm} M_{IFS} M_{Rot}(-\theta)
    M_{WP}(\phi) M_{Rot}(\theta) M_{AO} M_{tertiary} M_{telescope}  M_{SkyRot}(\eta) 
\] 
where $M_{Woll\pm}$ describes the Wollaston prism (with two variants for the two output orthogonal polarizations$+$ and $-$), $M_{WP}(\phi)$ represents a waveplate with retardance $\phi$ at angle $\theta$, and so on. The term $M_{SkyRot}(\eta)$ accounts for sidereal rotation by the parallactic angle $\eta$. All of the above may be implicitly a function of wavelength $\lambda$.  The instrumental polarization of the telescope is given by $M_{telescope}$ for the primary and secondary, times $M_{tertiary}$ for the contribution of the tertiary fold flat only if GPI is attached to a side port on the Gemini instrument support structure. Once each individual $M$ is known, either by design or empirical measurement, the set of equations of condition may be derived for any given observation sequence.

The process of measurement (Eq. \ref{Eq-least-squares}) stated in compact matrix form is
\[
I' = M' S.
\]
The least squares estimate of $S$ corresponds to minimizing the norm
\[
||I' - M' S_{est}||^2_2,
\]
which can be found using the pseudo-inverse of the measurement matrix
\begin{equation}
S_{est} = (M'^T M')^{-1} M'^T I' .
\label{eq-pseudo-inv}
\end{equation}

To summarize the approach we have implemented as part of the GPI Data Reduction Pipeline:
For each raw image, the two interlaced sets of polarized spots are extracted and assembled to produce a pair of images in orthogonal linear polarizations, as described in \S \ref{data_reduction}.  An observing sequence yields many such pairs. These data are rotated and aligned to a common orientation, the equations of condition are computed, and the resulting set of linear equations is then solved using singular value decomposition via IDL's \texttt{svdc} and \texttt{svsol} routines (Often $M'^TM'$ is ill-conditioned and the use of the pseudo-inverse in Eq. \ref{eq-pseudo-inv} should be thought of as schematic only.)  Because the field of view projected onto the sky changes with the parallactic angle, causing not all images in a sequence to contribute to all pixels in the final mosaic, this procedure must be repeated separately for each lenslet position. Still, the entire computation can be performed in only a few seconds on modern computers. The result is a 3D Stokes datacube [$x$, $y$, (I,Q,U,V)] derived from the entire observation sequence.

\subsection{A Variant on Double Differencing for ADI Polarimetry}
\label{appendix:doubledifferencing}

One of the strengths of differential polarimetry as an observational technique is the ability to modulate the signal.  Such modulation is not generally possible with other differential techniques; for instance one cannot simply modulate wavelengths in an integral field spectrograph.  By swapping a polarized signal between different measurement channels, non-common-path biases between the two channels can be mitigated.  

Let us denote the two orthogonally polarized channels in a given exposure as $I^+$ and $I^-$. The traditional double differencing approach to measure a polarized signal $Q$ would be to calculate difference images $d_i$ as follows, assuming a modulation that swaps $Q$ between the two channels:
\begin{eqnarray}
   d_1 &=& I^+_1 + I^-_1 = +Q + \epsilon \\
   d_2 &=& I^+_2 + I^-_2 = -Q + \epsilon 
\end{eqnarray}
Note that the differences are contaminated by some systematic bias $\epsilon$ between the two channels. By forming the double difference, we can recover the value of $Q$ while cancelling that bias: 
\begin{eqnarray}
  d_1 - d_2 &=& (+Q + \epsilon) - (-Q + \epsilon) = 2 Q
\end{eqnarray}

But as previously noted, for GPI the continuous parallactic angle rotation for ADI means we cannot in general choose waveplate angles to swap polarizations.  Because we only split the polarizations after the lenslet array, many forms of non-common-path error are minimized, but there can still be systematic biases between the two channels. In particular for GPI 
the dominant impact is from detector artifacts (bad pixels and uncorrected flat field residuals) and data cube assembly systematics which affect the two channels differently. 

In order to recover the benefits of double differencing in this situation we proceed as follows. Given an observation sequence, we begin by calculating the sum and difference images of each pair:
\begin{eqnarray}
  s_i &=& I^+_i + I^-_i \\
  d_i &=& I^+_i - I^-_i 
\end{eqnarray}
We then calculate the median of all the differences, $\Delta =$ median($d_i$).   This is our estimate of the systematic bias $\epsilon$ between channels, so we subtract it from each of the original differences to get a debiased difference, $d'_i$.   After doing so, we recombine the debiased differences with the sums $s_i$ to create debiased versions of the measured orthogonal polarizations $I'$:
\begin{eqnarray*}
  d'_i &=& d_i - \Delta\\
  I_i^{+'} &=&   (s_i + d'_i)/2\\
  I_i^{-'} &=&   (s_i - d'_i)/2
\end{eqnarray*}

These debiased orthogonal polarized intensities $I'$ are then used in place of the original measured polarized intensities when solving Equation~\ref{eq-pseudo-inv}.  Admittedly the median difference image, $\Delta$, is an imperfect estimator for the systematic offset between channels, and can itself be biased depending on the presence of polarized signals in the individual difference images, so the cancellation will not be perfect. However, given that we are mostly concerned with detector calibration and cube assembly artifacts that do not change on short timescales, the median computed over many images is relatively robust.
It is still worth pursuing improvements in calibration and algorithms to minimize the systematic biases between channels at their source, but empirically applying this correction is an efficient way to improve image quality and contrast in ADI polarimetry.

\subsection{Noise progagation in GPI Stokes Datacubes}
\label{appendix:noise}

An estimate of the noise in a pixel of a final Stokes cube can be found by propagating errors through Equation~\ref{eq-pseudo-inv} using standard statistical techniques. If the covariance matrix of the measured intensities $I'$ is $C_{I'}$, then the covariance of the estimate of the Stokes parameters $S_{est}$ is 
\begin{equation}
C_{S_{est}}= M^+ C_{I'} (M^+)^T, 
\end{equation}
where $m^+=(M'^T M')^{-1} M'^T$ is the pseudo-inverse of the measurement matrix. If the noise estimate in each element of $I'$ is identical (as is  the case when considering read-out noise or Poisson noise for an unpolarized source) then the covariance matrix $C_{I'}$ becomes a diagonal matrix with identical entries. The Stokes covariance matrix can then be expressed as: 
\begin{equation}
C_{S_{est}} = \sigma_{I'}^2 M^+(M^+)^T,
\end{equation}
where $\sigma_{I'}$ is an estimate of the noise in a single pixel of a polarimetry data cube.  

The read-out noise was calculated as $\sigma_{RON}=\sigma_{Det}*\sqrt{n_{pixels}/n_{coadds}}$, where $\sigma_{Det}$ is the read-out noise for a single detector pixel. For each annulus the Poisson noise variance in a polarization data cube pixel was taken to be half of the mean intensity of a Stokes $I$ cube in that annulus. 

If $Q$ and $U$ are assumed to be independent normally distributed quatities with means $\mu\cos{\theta}$ and $\mu\sin{\theta}$ and identical standard deviations $\sigma$, then the polarized intensity, $P$, can be described by a Rician distribution, $R(\mu,\sigma)$. For an unpolarized source (i.e. $\mu = 0$) then the RMS value of $P$ is equal to $2\sigma$. 

\vskip 1 cm

\bibliographystyle{apj}   
\bibliography{./references,./gpiSPIEbib}

\end{document}